%% file: main.tex
\documentclass[aps,pre,reprint,twocolumn,floatfix,nofootinbib]{revtex4-2}
\usepackage{siunitx}
\usepackage{graphicx}
\usepackage{physics}
\usepackage{amsmath}
\usepackage{bm}
\usepackage{bbm}
\usepackage{isotope}
\usepackage[ruled,vlined,linesnumbered]{algorithm2e}

\everymath{\displaystyle} % Use this to not get the ugly short integral symbols

\begin{document}
    \title{Exact weight cancellation in Monte Carlo eigenvalue transport problems}
    \author{Hunter Belanger}
    \email{hunter.belanger@cea.fr}
    \author{Davide Mancusi}
    \email{davide.mancusi@cea.fr}
    \author{Andrea Zoia}
    \email{andrea.zoia@cea.fr}
    \affiliation{Universit\'e Paris-Saclay, CEA, Service d'\'Etudes des R\'eacteurs et de Math\'ematiques Appliqu\'ees, 91191, Gif-sur-Yvette, France}
    
    \begin{abstract}
        Random walks are frequently used as a model for very diverse physical phenomena. The Monte Carlo method is a versatile tool for the study of the properties of systems modelled as random walks. Often, each walker is associated with a statistical weight, used in the estimation of observable quantities. Weights are typically assumed to be positive; nonetheless, some applications require the use of positive and negative weights or complex weights, and often pose particular challenges with convergence. In this paper, we examine such a case from the field of nuclear reactor physics, where the negative particle weights prevent the power iteration algorithm from converging on the sought fundamental eigenstate of the Boltzmann transport equation. We demonstrate how the use of weight cancellation allows convergence on the physical eigenstate. To this end, we develop a novel method to perform weight cancellation in an exact manner, in three spatial dimensions. The viability of this algorithm is then demonstrated on a reactor physics problem.
    \end{abstract}
    
    \maketitle
    
    \section{Introduction}
    Many physical processes can be be represented by the random movement of particles, or ``walkers'', through phase space. Such phenomena include radiation transport, propagation of active molecules in living bodies, or the spread of epidemics \cite{berg1993random,bartlett1960stochastic,martin1974,modest2013radiative}. Quite often, it is desirable to compute the properties of such systems with the aid of Monte Carlo simulations. This is certainly the case in neutron transport, where Monte Carlo techniques are used to sample the random walk process and thus solve the Boltzmann neutron transport equation \cite{bell_glasstone}. In this context, the Monte Carlo method may be preferred to (faster) deterministic methods because it requires very few (if any) approximations: Monte Carlo simulations are typically used to obtain reference solutions to which deterministic solutions may be compared.
    
    The random walkers have an associated statistical ``weight'', which is typically positive and real. This weight is used to estimate the observable quantities which are sought in the simulation. Certain forms of transport problems, however, require the use of positive and negative weights, or even complex weights. Some examples include diffusion quantum Monte Carlo \cite{arnow1982green}, the solution of neutron noise equations in the frequency domain \cite{yamamoto2013}, the determination of the second harmonic of the Boltzmann eigenvalue equation \cite{booth2003computing}, and the determination of transmittance in graphics rendering \cite{szirmay2017}. Monte Carlo simulations that use negative or complex weights are notoriously difficult, as they often do not converge to the desired solution \cite{arnow1982green, booth2003computing}, or have a very high variance in the observed quantities \cite{yamamoto2013}. Most literature recognizes that applying weight cancellation, where walkers carrying positive and negative weights may annihilate with one another, is highly beneficial for the solution of these problems \cite{BoothGubernatis2009,booth2003computing,booth2010exact,yamamoto2013,arnow1982green}.
    
    Several methods for performing weight cancellation have been proposed. One method, stemming from neutron transport, achieves weight cancellation by pairing walkers individually (as opposed to pooling weights of walkers within a defined region) \cite{booth2003computing}. While in theory this algorithm is exact (i.e.\ it does not introduce any approximation), it is difficult to implement, and has a quadratic computational complexity in the number of random walkers being simulated. Another linear, exact regional cancellation algorithm has been proposed \cite{booth2010exact}, but is only valid in 1D geometries. One alternative approach consists of using an approximate technique of averaging the weight of all walkers within the same region \cite{yamamoto2013}. This is easily applied to 3D, and has linear complexity, but does not provide an exact solution. Finally, it must be mentioned that not all random walker problems are posed in a manner which makes cancellation difficult. In the full configuration interaction quantum Monte Carlo technique, for example, cancellation is straightforward, as walkers explore a discrete state space, as opposed to a continuous state space; when two walkers of different signs land in the same state, they may immediately annihilate \cite{spencer2012}.
    
    In this paper, we review some of the existing weight cancellation methods and propose a novel exact 3D regional cancellation algorithm, that has linear complexity in the number of walkers, and that can be easily generalized to any number of dimensions. The development of this method has been motivated by certain neutron transport problems that require the use of negative walker weights in conjunction with the power iteration technique (for obtaining the dominant eigenstate), and that fail to converge to the correct solution without weight cancellation; a simplified example of such a system is presented in Section~\ref{sec:pid}. Subsequently, in Section~\ref{sec:theory}, we will develop a mathematical model to explain why the power iteration fails in the presence of negative weights, and why weight cancellation can resolve this problem. We will demonstrate how the existing 1D regional cancellation algorithm allows successful convergence of our 1D power iteration problem in Section~\ref{sec:1d_cancel}, and we will develop our novel, 3D version of the cancellation algorithm in Section~\ref{sec:3d_cancel}, testing it on a reactor physics benchmark in Section~\ref{sec:results}. Finally, we will present our conclusions Section~\ref{sec:conclusions}.
    
    \section{Power Iteration Debacle}\label{sec:pid}
    
     In neutron transport, a main quantity of interest is the fundamental eigenpair of the Boltzmann $k$-eigenvalue equation for neutrons, which we write here in its one-speed form, as a function of position $\bm{r}$ and direction $\hat{\bm{\Omega}}$:
     \begin{eqnarray}
        \bm{\hat{\Omega}}\cdot\nabla\varphi_k(\bm{r},\hat{\bm{\Omega}}) + \Sigma_t(\bm{r})\varphi_k(\bm{r},\hat{\bm{\Omega}}) = \nonumber\\ \int_{4\pi}\Sigma_s(\bm{r}, \hat{\bm{\Omega}}'\rightarrow\hat{\bm{\Omega}})\varphi_k(\bm{r},\hat{\bm{\Omega}}')\dd\hat{\bm{\Omega}}' + \nonumber\\ \frac{\nu(\bm{r})\Sigma_f(\bm{r})}{4\pi k} \int_{4\pi} \varphi_k(\bm{r},\hat{\bm{\Omega}}')\dd\hat{\bm{\Omega}}'.
        \label{eq:boltzmann}
    \end{eqnarray}
    Here $\varphi_k$ is the eigenfunction, $k$ is the eigenvalue, $\Sigma_t$ is the total cross section, $\Sigma_s$ is the scattering cross section, $\Sigma_f$ is the fission cross section, and $\nu$ is the average number of new particles produced per fission. While it does not explicitly appear in Eq.~\eqref{eq:boltzmann}, there is an implicitly defined capture cross section $\Sigma_c=\Sigma_t-\Sigma_s - \Sigma_f$, which results in the death of neutrons. In general, Eq.~\eqref{eq:boltzmann} admits several eigenvalue-eigenfunction pairs; the fundamental (largest) $k$-eigenvalue is the multiplication factor, which is typically written as $k_\text{eff}$. The corresponding eigenfunction $\varphi_0$ is known as the angular neutron flux, and it represents the average number of neutrons crossing a unit surface area per unit time \cite{bell_glasstone}.
    
    In order to obtain $k_\text{eff}$ and $\varphi_0$, a numerical technique known as power iteration is often employed \cite{bell_glasstone,holmes2016introduction}. Although Eq.~\eqref{eq:boltzmann} is not written in such a form, we may assume that it represents an eigenvalue problem of the type $Lv_0 = \lambda_0 v_0$, with $v_0$ being the fundamental eigenstate, and $\lambda_0$ being the fundamental eigenvalue. In relation to Monte Carlo transport problems, the operator $L$ can be interpreted as the propagation of particles through a system from one fission event to another. With power iteration, one may obtain the fundamental state $v_0$ from any state $b$ such that the inner product $\langle b, v_0\rangle \not = 0$, by repeated application of the operator $L$:
    \begin{equation}
        \lim_{n\rightarrow\infty} \frac{L^nb}{\abs{L^nb}} = v_0.
        \label{eq:power_it}
    \end{equation}
    Therefore, starting from almost any initial state, it is possible to converge to the fundamental mode \cite{holmes2016introduction}.
    
    We wish to use Monte Carlo to yield a solution to Eq.~\eqref{eq:boltzmann}, in the sense that the average density of walkers in phase space should satisfy Eq.~\eqref{eq:boltzmann}, and represent the fundamental eigenmode. In the context of Monte Carlo, the idea of the power iteration technique needs to be adapted as follows. Starting from an arbitrary set of walkers, we sample random walks for all walkers until their death. Along the walk, a walker may be randomly killed if the magnitude of its weight becomes too low, or split if the magnitude becomes too large \cite{lux}. During the random walks of this first generation, new particles will be born from fission. These fission particles are stored in a bank and are attributed to the second generation. Once the first generation has finished, the banked particles undergo population control, where those with small weights may randomly be killed, those with large weights may be split into multiple particles, and the net weight of all particles is normalised. This is done to keep the net weight of all the particles at the beginning of a generation constant, and to keep the number of particles in the simulation constant on average \cite{lux}. This culled and normalized particle bank is subsequently used as the source for the second generation. These particles then undergo the random walk, producing fission particles which will belong to the third generation. This application of the random walk mechanics on a generation of particles may be continued indefinitely. After a number of a generations, the positions (and directions) of the fission particles will settle on an equilibrium distribution, representing the converged fission source for the problem \cite{lux}. The combined random walks of all particles from this converged fission source represent the dominant eigenfunction of Eq.~\eqref{eq:boltzmann}. With the fission source converged, observable quantities (such as the angular flux and the multiplication factor) may be estimated in each generation. With an estimation of the angular flux and the multiplication factor obtained by each generation, an average may be calculated, although the estimation of the uncertainty is not trivial \cite{nowak2016}. 
    
    Within a generation, the random walk process is sampled as follows. A particle begins with an initial position and direction, which for the first generation may be sampled from a somewhat arbitrary distribution. The first task is to sample the distance the particle will fly before having a collision. In most applications of Monte Carlo simulation to reactor physics problems, the total cross section is usually assumed to be piece-wise constant within each macroscopic geometric region composing the modeled system. To sample a flight distance in a material region with a spatially constant cross section, a random variable $\xi\sim\mathcal{U}[0,1)$ is drawn, and the cumulative distribution function for the flight distance must be inverted to obtain the distance to collision $d_c$\footnote{It is for this reason that the underlying random walk is called ``exponential flight'' \cite{zoia2011}.}: 
    \begin{equation}
        \xi = \int_0^{d_c} \Sigma_t \exp(-\Sigma_t s) \dd s \Rightarrow d_c = -\frac{\ln(1-\xi)}{\Sigma_t}.
        \label{eq:constant_sample_d}
    \end{equation}
    The sampled distance is only valid in the given material region. If the distance $d_m$ to the next material boundary is less than $d_c$, the particle is only moved by $d_m$, the cross section is updated, and a new distance to collision is sampled. Once at a collision site, the collision mechanics may be simulated; the particle can be captured with probability $\Sigma_c/\Sigma_t$, whereupon the history is terminated; alternatively, scattering events may occur with probability $\Sigma_s/\Sigma_t$, or the particle may produce new fission particles with probability $\Sigma_f/\Sigma_t$, which will be part of the next generation \cite{lux}. The multiplication factor may be estimated in several different manners: here we use the collision estimator. When a particle of weight $w$ has a collision, we add the expected number of fission neutrons produced per collision at that location, namely $w\nu(\bm{r})\Sigma_f(\bm{r})/\Sigma_t(\bm{r})$, to an accumulator which we shall refer to as $K$. Once the generation has finished, the estimate for the multiplication factor for the generation $G$ is
    \begin{equation}
        k_\text{eff}^{(G)} = \frac{K}{W},
    \end{equation}
    $G$ being the generation number, and $W$ being the sum of the weights of all the particles at the beginning of the generation (which is a constant, because of the normalization procedure described above). The accumulator $K$ must of course be reset to zero before each new generation begins.
    
    A more realistic description of the system would consist of relaxing the hypothesis of piece-wise constant cross sections, especially in view of multi-physics problems \cite{demaziere2013} where the cross sections for neutron transport depend on complex space-dependent physical feedback mechanisms such as temperature and material density fields. Recently, we examined different methods of sampling the flight distance for the case of spatially varying cross sections \cite{belanger2020review}. Such methodologies are highly desirable for the next generation of Monte Carlo transport codes, as they could allow for a better representation of the system being simulated, reduce memory requirements, and pair well with the reactor physics community's goals of taking into account a multi-physics approach \cite{demaziere2019modelling}. In principle, sampling the distance to collision for spatially dependent cross sections would require inverting the following equation for $d_c$:
    \begin{equation}
        \xi = \int_0^{d_c}\Sigma_t(s\hat{\bm{\Omega}} + \bm{r}_0)\exp(\int_0^s\Sigma_t(u\hat{\bm{\Omega}}+\bm{r}_0)\dd u)\dd s.
        \label{eq:cont_fd}
    \end{equation}
    In practice, this is quite difficult to accomplish. Our previous work considered several algorithms which could sample flight distances from spatially continuous cross sections, without needing to directly invert Eq.~\eqref{eq:cont_fd}. In particular, we examined the traditional delta tracking \cite{woodcock1965techniques,leppanen2010performance}, and the negative weighted delta tracking methods \cite{legrady2017woodcock,molnar2018variance}. 
    
    In order to illustrate how these algorithms work, we revisit a simple one-dimensional transport problem that was considered in our previous paper. In the so-called `rod model', particles may only move in the forward or backward direction, along a line segment. For our purposes, this line segment has a finite length, and it is possible for particles to leak out of either end of the line. For our application we will set the boundaries of the segment at $x=0$ and $x=2$. As an example of a space-dependent cross section, we will use the broad-Gaussian cross section from our earlier paper \cite{belanger2020review}. This cross section profile was chosen as it is not monotonic, and better represents the idea that particles may see both an increase and decrease in the cross section along their ray of flight. The chosen $\Sigma_t(x)$ has the form
    \begin{equation}
        \Sigma_t(x) = \sqrt{\frac{2}{\pi}}e^{-(x-1.23)^2} + 0.1 \text{ }\forall x\in [0,2]\text.
    \end{equation}
    The scattering, absorption and fission probabilities are chosen to be spatially constant, having values of $\Sigma_s/\Sigma_t=0.7$, $\Sigma_c/\Sigma_t=0.1$, and $\Sigma_f/\Sigma_t=0.2$. The average number of neutrons born per fission is $\nu=2.5$; both scattering and fission are isotropic (equal probability of emission forward or backward).
    
    \begin{algorithm}[t]
        \SetAlgoLined
        Sample uniform random variable $\xi_1$\;
        $d := -\ln(\xi_1)/\Sigma_\text{smp}$\;
        $\bm{r}_1 := d\hat{\bm{\Omega}} + \bm{r}_0$\;
        Sample random variable $\xi_2$\;
        \eIf{$\xi_2 < q $}{
            $\displaystyle w := 
            w\frac{\Sigma_t(\bm{r}_1)}{q\,\Sigma_\text{smp}}$\;
            Perform real collision\;
        }{
            $\displaystyle w := w \frac{1 - 
            \frac{\Sigma_t(\bm{r}_1)}{\Sigma_\text{smp}}}{1-q}$\;
            Virtual collision; goto line 1\;
        }
        \caption{Negative weighted delta tracking}
        \label{alg:ndwt}
    \end{algorithm}
    
    Transport is conducted using the negative weighted delta tracking (NWDT) method \cite{legrady2017woodcock,molnar2018variance}. The basic procedures for this method are presented in Alg.~\ref{alg:ndwt}. It relies on the concept of real and virtual collisions. NWDT requires two parameters: a sampling cross section $\Sigma_\text{smp}$, which is used to sample the distance to a tentative collision site, and a probability $q$, which is used to determine whether a collision is real or virtual. At a real collision, the mechanics of a collision are used to change the direction of the particle accordingly. A virtual collision does not simulate the collision mechanics; instead, the direction of the particle is left unchanged, and the distance to a new tentative collision site is sampled, where the process begins again. Both $\Sigma_\text{smp}$ and $q$ are allowed to be functions of the position. According to the NWDT algorithm, the particle weight changes sign when a virtual collision is sampled, and $\Sigma_t > \Sigma_\text{smp}$. Here, we have chosen $\Sigma_\text{smp} = 0.85\Sigma_t(x=1.23)$, which leads to particles changing sign within the region $0.8\lesssim x \lesssim 1.66$. As for $q$, we have used $q=\Sigma_t/(\Sigma_t - \abs{\Sigma_\text{smp}-\Sigma_t})$, which corresponds to the strategy proposed by Carter, Cashwell and Taylor \cite{carter1972monte,molnar2018variance}. This choice of $q$ was motivated by our previous work, where the method demonstrated very reasonable performance compared to the standard of delta tracking \cite{belanger2020review}.
    
    Delta tracking is a special case of NWDT, where the sampling cross section $\Sigma_\text{smp}$ is taken to be a strict majorant cross section $\Sigma_\text{maj}$, with $\Sigma_\text{maj}\geq\Sigma_t$ and the acceptance probability $q$ is taken equal to $\Sigma_t/\Sigma_\text{maj}$ \cite{woodcock1965techniques,leppanen2010performance}\footnote{Delta tracking is also sometimes referred to as self-scattering in the electron transport community \cite{jacoboni1983}, as thinning \cite{lemaire2018pdm}, or as the null-collisions method \cite{galtier2013}.}. With these choices, particle weights never change sign; however, insisting on the use of delta tracking can raise a number of practical problems, because it could be very challenging to obtain a majorant when using spatially continuous cross sections in a more realistic context. If one were to conduct a simulation where both the isotopic density and the temperature of materials varied spatially, there is, in general, no way to exactly determine the majorant cross section for the system. It is certainly possible to probe the phase space of the problem, testing the cross section at each point. However, if the selected sampling cross section is not in fact a majorant, it will be impossible to handle collisions for which $\Sigma_t>\Sigma_\text{smp}$; even if such an event does not occur, the results will be slightly biased. Conversely, if $\Sigma_\text{smp}$ is chosen by applying a large safety margin to the maximum known cross section (in an effort to ensure underestimation of the majorant does not occur), the method becomes quite inefficient, as many unnecessary virtual collisions will occur \cite{leppanen2010performance}. NWDT tolerates underestimations in the majorant cross section, and also allows one to avoid gross overestimations of the majorant which reduce efficiency.
    
    With NWDT, not only may there be negatively weighted particles in the system, but the weights may also change in magnitude. These two traits can lead to an increase in variance and simulation time. To mitigate these effects, roulette is used on particles with $\abs{w} < 0.6$, with the survival weight being $w=\pm1$ (ensuring that the particle keeps its initial sign). Particles are also split if $\abs{w} \ge 2$.
    
    In our previous work, we neglected fission. In view of testing NWDT in the framework of power iteration for eigenvalue problems, fission has been added to our model. The number $n$ of new fission neutrons generated at any collision is taken to be
    \begin{equation}
        n = \left\lfloor\abs{w}\frac{\nu\Sigma_f}{\Sigma_t}\cdot\frac{1}{k_\text{eff}^{(G-1)}} + \xi\right\rfloor,
        \label{eq:n_keff}
    \end{equation}
    where $\xi\sim \mathcal{U}[0,1)$ is a uniform random variable, and $k_\text{eff}^{(G-1)}$ is the estimated value of $k_\text{eff}$ for the previous generation of particles. Without dividing by $k_\text{eff}^{(G-1)}$, the number of particles in the simulation would either increase exponentially if the system is super-critical ($k_\text{eff} > 1$), or decrease exponentially if the  system is sub-critical ($k_\text{eff} < 1$). This algorithm closely follows standard methods applied in Monte Carlo codes \cite{mcnpTheory, openmc}, with slight modifications to accommodate fission with negative particles. All new fission particles are born with a weight $w=\pm1$, keeping the sign of the weight of the particle inducing fission. Between particle generations, all particle weights are multiplied by a normalization coefficient to ensure that the net weight of the system (sum of all particle weights) is always a constant value.
    
    \begin{figure}
        \includegraphics[width=\columnwidth]{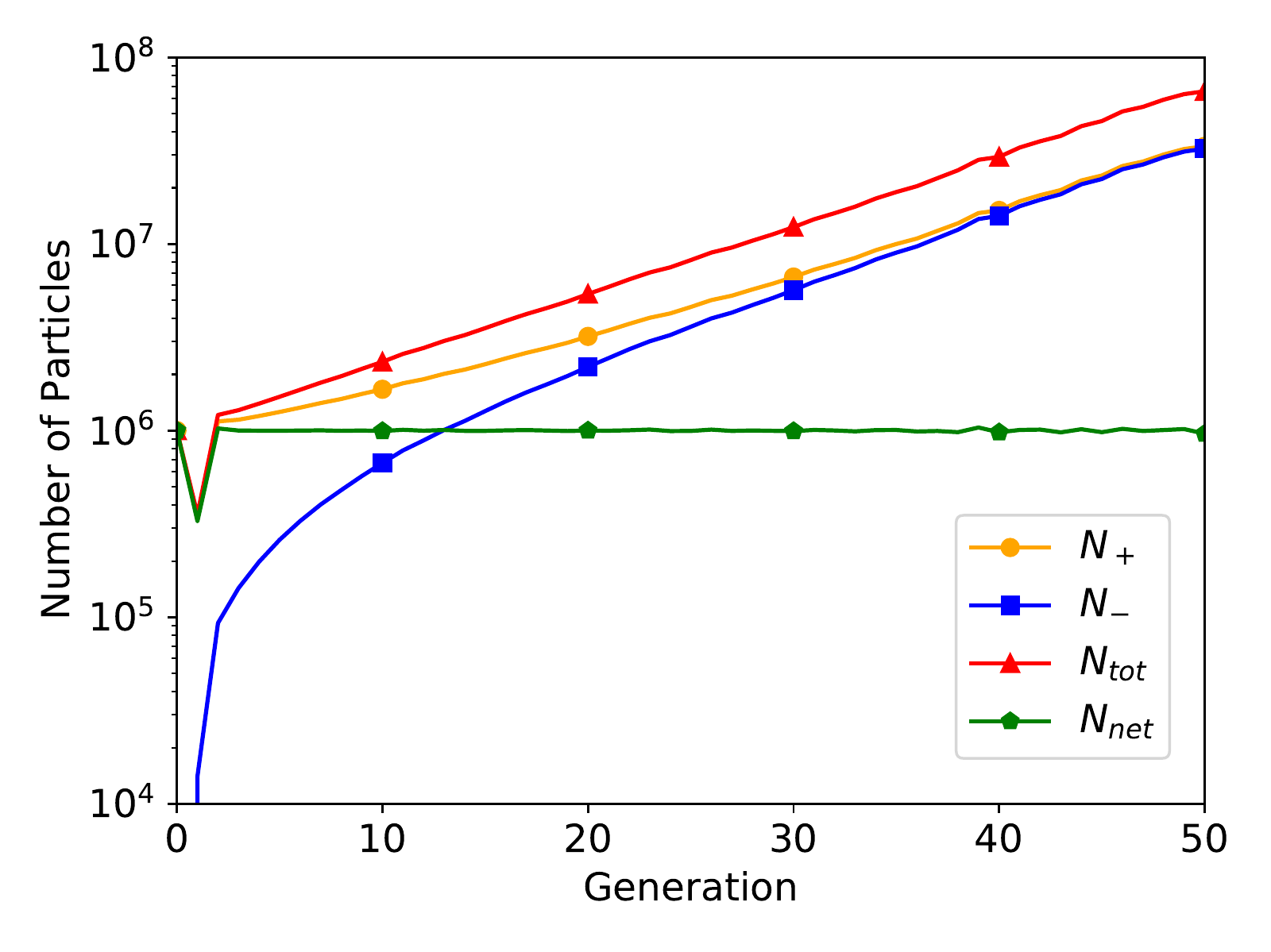}
        \caption{The number of positive and negative particles per generation, when using the population control scheme provided in Eq.~\eqref{eq:n_keff}.}
        \label{fig:1d_keff_nc}
    \end{figure}
    
    \begin{figure}
        \includegraphics[width=\columnwidth]{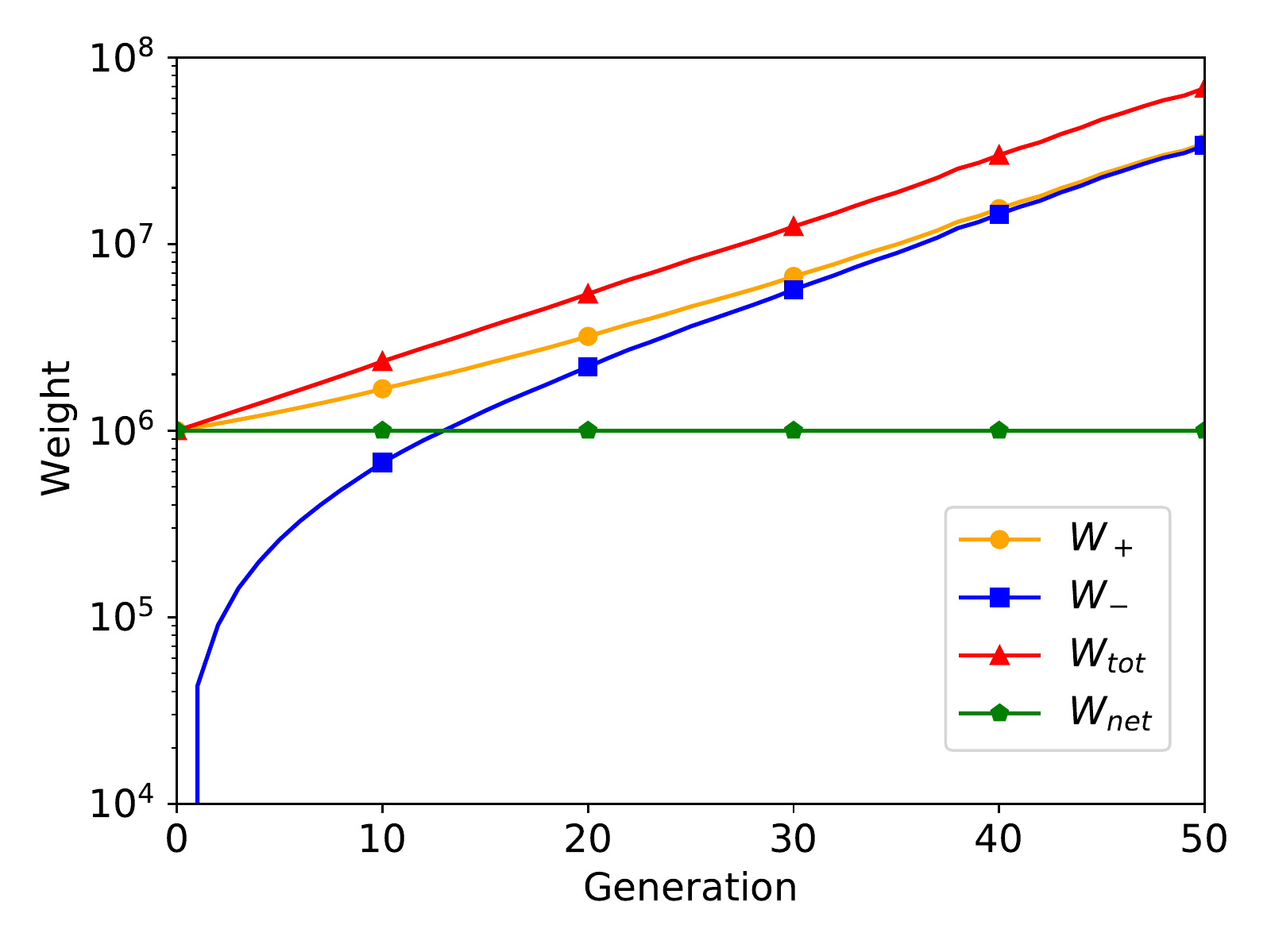}
        \caption{The weight of positive and negative particles per generation, when using the population control scheme provided in Eq.~\eqref{eq:n_keff}.}
        \label{fig:1d_keff_wc}
    \end{figure}
    
    Figure~\ref{fig:1d_keff_nc} shows the number of positive and negative particles per generation, when starting with $10^6$ particles uniformly distributed within the rod. Also shown is the net number of particles $N_\text{net} = N_+ - N_-$, and the total number of particles $N_\text{tot}=N_+ + N_-$. Similarly, Figure~\ref{fig:1d_keff_wc} shows the total positive weight $W_+$ (the sum of the weights of all positive particles), the total negative weight $W_-$ (the sum of the magnitude of the weights of all negative particles), together with the net weight $W_\text{net} = W_+ - W_-$ and the total weight $W_\text{tot} = W_+ + W_-$. The number of particles is somewhat less meaningful than the weight, as one can use methods such as weight combing \cite{booth1996weight} to adjust how many particles are followed in the simulation, but this says nothing about the magnitude of their weight. Weight is the more natural quantity to examine, as its behavior is not controlled by the number of particles. It is instructive, however, to consider the interplay between weight and number of particles. While Eq.~\eqref{eq:n_keff} keeps the {\em net} number of particles stable as expected, the {\em total} number of particles is left to increase exponentially. Eventually, this increase in the number of particles causes each generation to take a longer amount of time to process, and the memory requirements increase as well. At some point, the computer memory is overwhelmed, and the simulation fails. In an attempt to remedy this problem, we modified Eq.~\eqref{eq:n_keff} by replacing $k_\text{eff}^{(G-1)}$; using the value of $k_\text{eff}$ for the previous generation keeps the {\em net} number of particles constant on average, so a new quantity $k_\text{tot}$ was defined, representing the the increase in the {\em total} weight $W_\text{tot}$ of the simulation. The quantity $k_\text{tot}$ may be estimated in a similar manner to the multiplication factor, accumulating the collision estimator $\abs{w}\nu(\bm{r})\Sigma_f(\bm{r})/\Sigma_t(\bm{r})$. Using $k_\text{tot}^{(G-1)}$ in Eq.~\eqref{eq:n_keff}, we obtain
    \begin{equation}
        n = \left\lfloor\abs{w}\frac{\nu\Sigma_f}{\Sigma_t}\cdot\frac{1}{k_\text{tot}^{(G-1)}} + \xi\right\rfloor.
        \label{eq:n_ktot}
    \end{equation}
    
    \begin{figure}
        \includegraphics[width=\columnwidth]{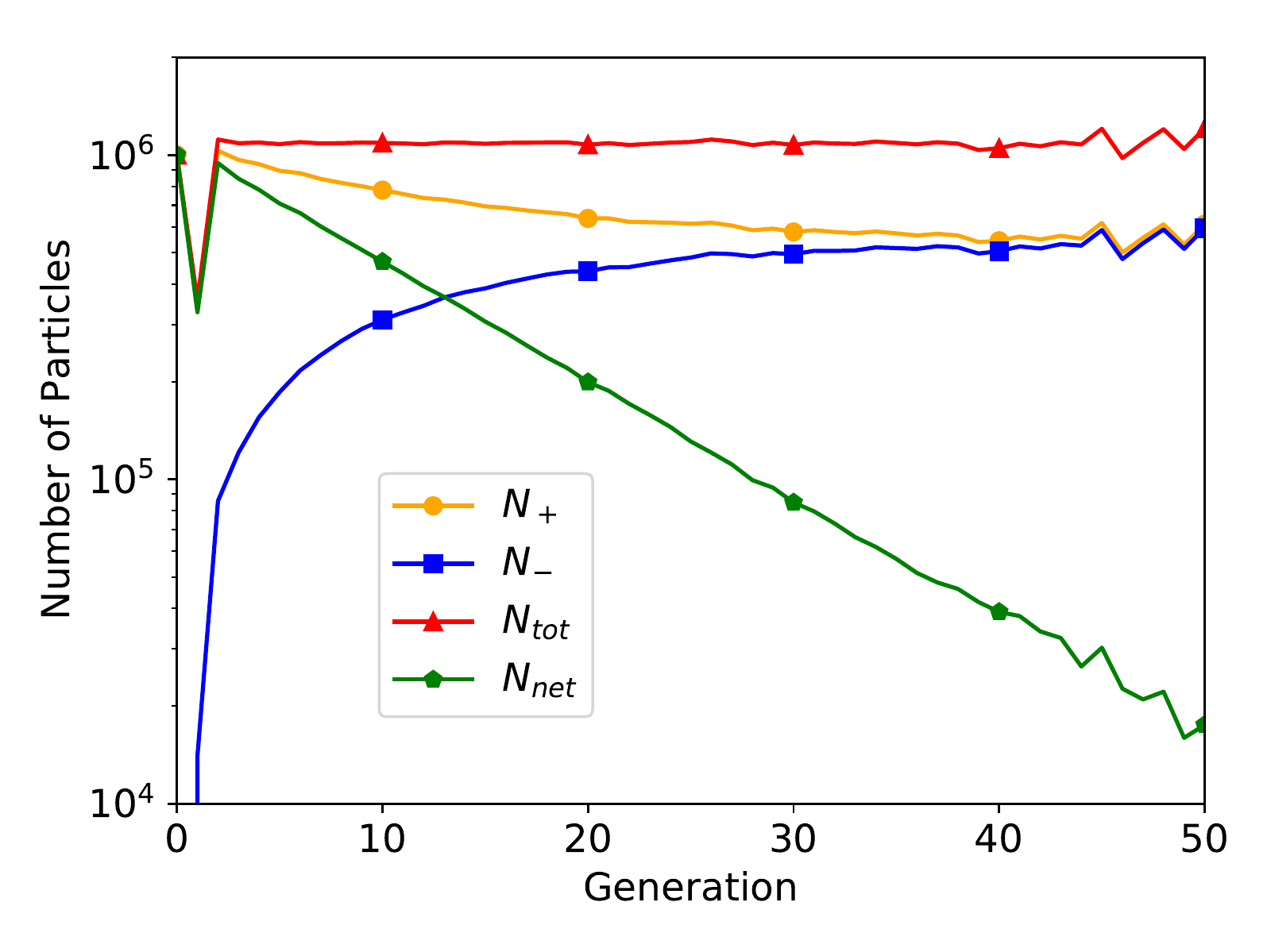}
        \caption{The number of positive and negative particles per generation, when using the population control scheme provided in Eq.~\eqref{eq:n_ktot}.}
        \label{fig:1d_ktot_nc}
    \end{figure}
    
    \begin{figure}
        \includegraphics[width=\columnwidth]{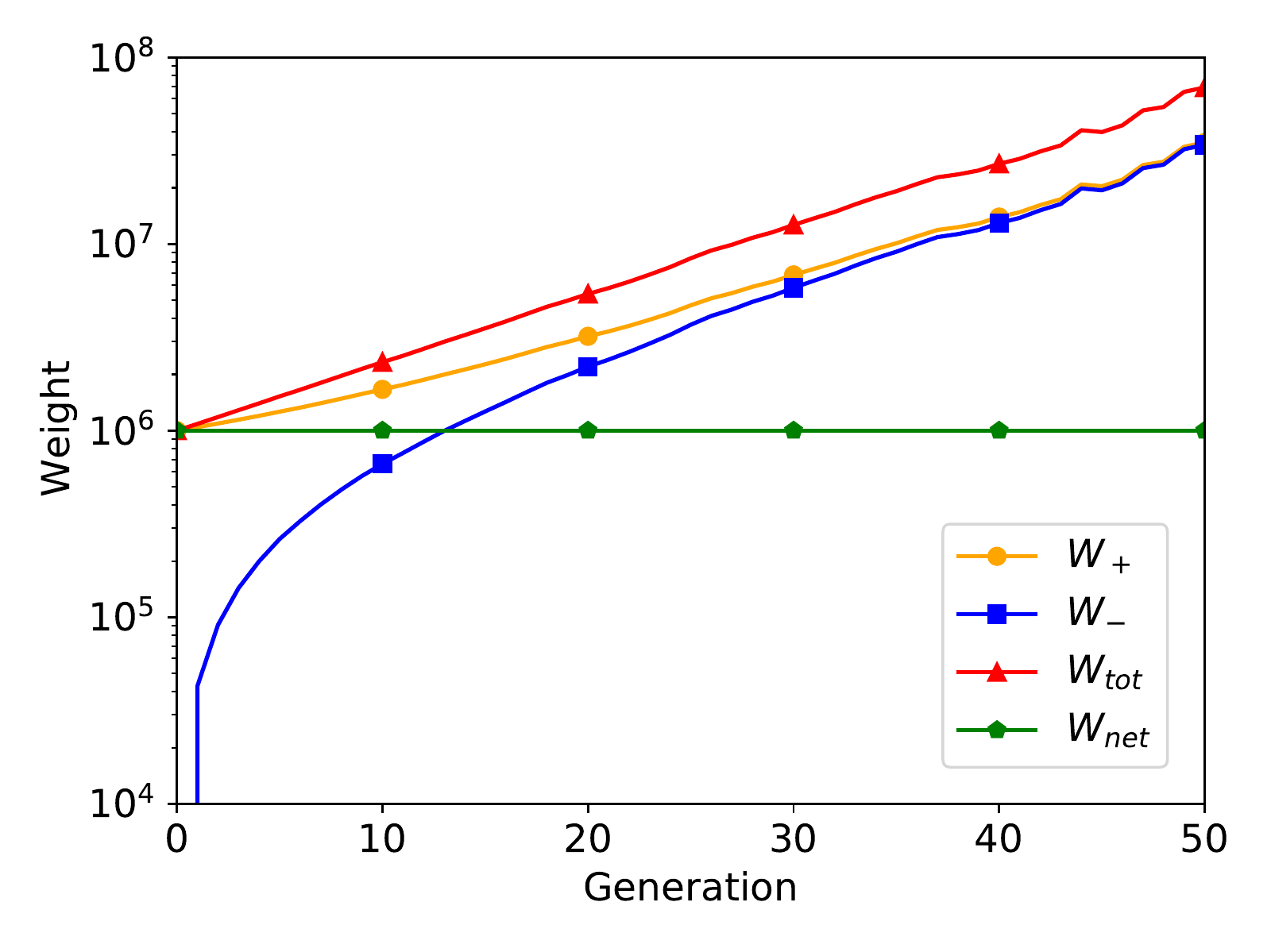}
        \caption{The weight of positive and negative particles per generation, when using the population control scheme provided in Eq.~\eqref{eq:n_ktot}.}
        \label{fig:1d_ktot_wc}
    \end{figure}
    
    The effects on the particle populations from using this normalization technique are presented in Figure~\ref{fig:1d_ktot_nc} and Figure~\ref{fig:1d_ktot_wc}. While the total number of particles indeed remains constant on average, the net number of particles now decreases exponentially. The net weight of course remains constant because we are normalizing it at the end of the generation. With fewer net particles to maintain the entire initial weight of the system, the magnitude of the weight of the particles increases drastically. This is indicated by the total weight of the system continuing to increase exponentially in Figure~\ref{fig:1d_ktot_wc}. This leads to near equal quantities of positive and negative particles, each with very large weight magnitudes, causing large fluctuations in the flux estimators, and therefore a larger variance in scores.
    
    Other population control mechanisms were tested as well, hoping that they might stabilize the system. One such method was particle combing \cite{booth1996weight}, conducted between each generation of particles. While combing did reduce the total number of particles going into each generation, the weight magnitude of each particle was quite large. Once they entered the simulation, the particles were split, causing an abrupt increase in memory consumption. After several generations, the simulation is killed by the operating system due to the large memory consumption.
    
    Although the rod model example is admittedly simple, similar results have been obtained when using the NWDT algorithm in combination with power iteration on more realistic reactor physics problems in 3D, and with energy-dependent cross sections (see Sec.~\ref{sec:results}). It is clear that the power iteration with NWDT cannot be used to estimate the equilibrium distribution of walkers, and therefore the solution to Eq.~\eqref{eq:boltzmann}. It is therefore pertinent to better understand why the power iteration is failing, and how the problem might be addressed.
    
    \section{Necessity of Weight Cancellation}\label{sec:theory}
    
    \subsection{The failure of power iteration}
    The Boltzmann equation presented in Eq.~\eqref{eq:boltzmann} is not capable of describing the interaction between the positive and negative particle populations in a simulation, only their combined average angular flux. It is fairly straightforward, however, to propose a formulation of the transport equation which is capable of representing both positively and negatively weighted particles. Considering Alg.~\ref{alg:ndwt}, outlining the NWDT method, we may write a set of coupled transport equations, one for the angular flux $\varphi_+$ of the positive particles, and a second for the angular flux $\varphi_-$ of the negative particles, resulting in the following coupled equations:
    \begin{eqnarray}
        \hat{\bm{\Omega}}\cdot\nabla\varphi_\pm + \Sigma_\text{smp}\varphi_\pm = \mathcal{S}\varphi_\pm + \frac{1}{k}\mathcal{F}\varphi_\pm + \nonumber\\ \Delta(\Sigma_\text{smp}-\Sigma_t)\,\varphi_\pm + \Delta(\Sigma_t-\Sigma_\text{smp})\,\varphi_\mp.
        \label{eq:cpld}
    \end{eqnarray}
    Here, $\mathcal{S}$ is defined as
    \begin{eqnarray}
        \mathcal{S}\varphi = \int_{4\pi}\Sigma_s(\bm{r},\hat{\bm{\Omega}}'\rightarrow\hat{\bm{\Omega}})\varphi(\bm{r},\hat{\bm{\Omega}}')\dd\hat{\bm{\Omega}}'
        \label{eq:S_deff}
    \end{eqnarray}
    and $\mathcal{F}$ is defined as
    \begin{eqnarray}
        \mathcal{F}\varphi = \frac{\nu(\bm{r})\Sigma_f(\bm{r})}{4\pi}\int_{4\pi}\varphi(\bm{r},\hat{\bm{\Omega}}')\dd\hat{\bm{\Omega}}'.
        \label{eq:F_deff}
    \end{eqnarray}
    We have also made use of the function
    \begin{equation}
        \Delta(x) = \begin{cases} x & x > 0 \\ 0 & x \leq 0\end{cases}.
        \label{eq:diode}
    \end{equation}
     A rigorous derivation of this statement is given in the Appendix. There is a subtlety in this formulation: instead of having one species of particles which may have a weight that is positive or negative, we now have two species (positive and negative), and both are represented by particles having a strictly positive weight. In this framework, our unknown is now the set of eigenfunctions
    \begin{eqnarray}
        \bm{\zeta} = \begin{bmatrix} \varphi_+ \\ \varphi_- \end{bmatrix},
        \label{eq:zeta_deff}
    \end{eqnarray}
    which is the solution of the generalized eigenvalue problem
    \begin{equation}
        \bm{A}\bm{\zeta} = \frac{1}{k}\bm{F}\bm{\zeta},
        \label{eq:cpld_mtrx}
    \end{equation}
    operator $\bm{A}$ being
    \begin{equation}
        \bm{A} = \begin{bmatrix}
        \mathcal{A}_{11}&\mathcal{A}_{12}\\
        \mathcal{A}_{12}&\mathcal{A}_{11}
        \end{bmatrix}\text,
        \label{eq:A}
    \end{equation}
    with
    \begin{equation}
    \begin{gathered}
        \mathcal{A}_{11}=\hat{\bm{\Omega}}\cdot\nabla + \Sigma_\text{smp} - \mathcal{S} - \Delta(\Sigma_\text{smp}-\Sigma_t)\text, \\
        \mathcal{A}_{12}=-\Delta(\Sigma_t - \Sigma_\text{smp}),
    \end{gathered}\label{eq:A11_A12}
    \end{equation}
    %\begin{widetext}
    %\begin{eqnarray}
    %    \bm{A} = \begin{bmatrix}\hat{\bm{\Omega}}\cdot\nabla + \Sigma_\text{smp} - \mathcal{S} - \Delta(\Sigma_\text{smp}-\Sigma_t) && -\Delta(\Sigma_t - \Sigma_\text{smp}) \\ -\Delta(\Sigma_t - \Sigma_\text{smp}) && \hat{\bm{\Omega}}\cdot\nabla + \Sigma_\text{smp} - \mathcal{S} - \Delta(\Sigma_\text{smp}-\Sigma_t)\end{bmatrix}
    %    \label{eq:A}
    %\end{eqnarray}
    %\end{widetext}
    and $\bm{F}$ being defined as
    \begin{equation}
        \bm{F} = \begin{bmatrix}\mathcal{F} && 0 \\ 0 && \mathcal{F}\end{bmatrix}.
    \end{equation}
    If we use $\mathcal{V}$ to denote the vector space of the physical flux, then $\bm{\zeta}$ is an element of the expanded vector space $\mathcal{V}\times\mathcal{V}$.
    
    Given the previous definitions, the physical flux is interpreted to be
    \begin{equation}
        \varphi = \varphi_+ - \varphi_-
        \label{eq:phys_flux_from_weighted}
    \end{equation}
    and may be retrieved through the application of $\bm{D}:\mathcal{V}\times\mathcal{V}\rightarrow\mathcal{V}$, defined as
    \begin{equation}
        \bm{D}=\begin{bmatrix}I && -I\end{bmatrix}
    \end{equation}
    and whose action is
    \begin{equation}
        \bm{D}\begin{bmatrix}f \\ g\end{bmatrix} = f - g.
    \end{equation}
    
    The eigenvalue equation presented in Eq.~\eqref{eq:cpld_mtrx} has several interesting properties, which may be elegantly outlined by the introduction of the parity operator, namely
    \begin{equation}
        \bm{P} = \begin{bmatrix}0 && I \\ I && 0\end{bmatrix}.
    \end{equation}
    Eq.~\eqref{eq:cpld_mtrx} is invariant under the action of the parity operator: that is, if $\bar{\bm{\zeta}}$ is a solution of Eq.~\eqref{eq:cpld_mtrx} with eigenvalue $\bar{k}$, then so is $\bm{P}\bar{\bm{\zeta}}$. %This is demonstrated by starting with
    %\begin{equation}
    %    \bm{A}\bar{\bm{\zeta}} = \frac{1}{\bar{k}}\bm{F}\bar{\bm{\zeta}},
    %\end{equation}
    %then inserting $\bm{P}^2=\bm{I}$ on both sides, and finally multiplying %from the left by $\bm{P}$ on both sides:
    %\begin{equation}
    %    \bm{P}\bm{A}\bm{P}(\bm{P}\bar{\bm{\zeta}}) = %\frac{1}{\bar{k}}\bm{P}\bm{F}\bm{P}(\bm{P}\bar{\bm{\zeta}}).
    %\end{equation}
    It is easily verifiable that $\bm{P}\bm{A}\bm{P}=\bm{A}$ and $\bm{P}\bm{F}\bm{P}=\bm{F}$, and therefore
    \begin{equation}
        \bm{A}(\bm{P}\bar{\bm{\zeta}}) = \frac{1}{\bar{k}}\bm{F}(\bm{P}\bar{\bm{\zeta}}).
    \end{equation}
    Thus, if the eigenvalue $\bar{k}$ is non-degenerate, we must have $\bm{P}\bar{\bm{\zeta}}=\pm\bar{\bm{\zeta}}$. This partitions the eigenstates into two sets, depending on the sign of their eigenvalue with respect to $\bm{P}$. The odd eigenstates have the form
    \begin{equation}
        \bm{\zeta}_o = \begin{bmatrix}\varphi \\ -\varphi \end{bmatrix}.
    \end{equation}
    By inserting this ansatz into Eq.~\eqref{eq:cpld_mtrx}, we can verify that $\varphi$ must solve the physical eigenvalue equation
    \begin{equation}
        \bm{\hat{\Omega}}\cdot\nabla\varphi + \Sigma_{t}\varphi = \mathcal{S}\varphi + \frac{1}{k}\mathcal{F}\varphi,
        \label{eq:op_boltzmann}
    \end{equation}
    which is just Eq.~\eqref{eq:boltzmann}. We denote the eigenfunctions and eigenvalues of this equation as $\varphi_i$ and $k_{\varphi,i}$, respectively, with $i\in\{0,1,\ldots\}$. The even eigenstates have the form
    \begin{equation}
        \bm{\zeta}_e = \begin{bmatrix}\eta \\ \eta \end{bmatrix},
    \end{equation}
    where $\eta$ must solve the modified Boltzmann equation
    \begin{equation}
        \bm{\hat{\Omega}}\cdot\nabla\eta + \Sigma_{t,\eta}\eta = \mathcal{S}\eta + \frac{1}{k}\mathcal{F}\eta,
        \label{eq:modified_boltzmann}
    \end{equation}
    which is verified by substitution into Eq.~\eqref{eq:cpld_mtrx}. Here we have defined
    \begin{equation}
        \Sigma_{t,\eta}=\Sigma_\text{smp}-\abs{\Sigma_\text{smp}-\Sigma_t}.
    \end{equation}
    We denote the eigenfunctions and eigenvalues of Eq.~\eqref{eq:modified_boltzmann} as $\eta_i$ and $k_{\eta,i}$, respectively, with $i\in\{0,1,\ldots\}$.
    
    Since we always have $\Sigma_{t,\eta}\leq\Sigma_t$, the equation for $\eta$ has the same scattering and fission terms as Eq.~\eqref{eq:boltzmann}, but capture has been decreased, namely
    \begin{eqnarray}
       \Sigma_{c,\eta} = \Sigma_{t,\eta} - \Sigma_s - \Sigma_f \le \Sigma_t - \Sigma_s - \Sigma_f.
    \end{eqnarray}
    Thus, on physical grounds, we correspondingly expect a larger dominant eigenvalue:
    \begin{equation}
        k_{\eta,0}\geq k_{\varphi,0}=k_\text{eff}.
        \label{eq:kn_gt_keff}
    \end{equation}
    Therefore, the dominant eigenvalue of Eq.~\eqref{eq:cpld_mtrx} is not $k_\text{eff}$, but the nonphysical eigenvalue $k_{\eta,0}$. Applying the power iteration method to Eq.~\eqref{eq:cpld_mtrx} will result in convergence towards the latter eigenstate. Additionally, the equilibrium distribution is even under the exchange of positive and negative particles, i.e.\ it contains the same amount of positive and negative particles, and thus zero net particles. Converging to a state which has zero net particles is incompatible with traditional population control mechanisms (such as combing, or normalizing by $k_\text{eff}$), which are designed to keep the net weight constant, leading to the divergence in particle populations which was observed in Sec.~\ref{sec:pid}.
    
    It is worth stressing the similarity between our analysis and the study by Spencer et al.\ \cite{spencer2012} on the origin of the sign problem in full configuration interaction quantum Monte Carlo without weight cancellation.
    
    \subsection{Modeling weight cancellation}
    
    Previous investigations have shown that weight cancellation can be very effective in dealing with Monte Carlo problems using particles of both positive and negative weights \cite{rouchon2017,booth2003computing,arnow1982green,spencer2012}. In view of these considerations, in the following we will formally address the effect of cancellation on Eq.~\eqref{eq:cpld_mtrx}. Weight cancellation can be modeled in the following manner: we start with some function of the phase space variables $(\bm{r},\hat{\bm{\Omega}})$, i.e.\ an element of $\mathcal{V}$. This function may be embedded in $\mathcal{V}\times\mathcal{V}$ by applying a suitable mapping $\bm{E}:\mathcal{V}\rightarrow\mathcal{V}\times\mathcal{V}$. There is some latitude in the definition of $\bm{E}$; one possible definition is
    \begin{equation}
        \bm{E}_1f=\begin{bmatrix}f \\ 0\end{bmatrix},
    \end{equation}
    but equally valid choices could be
    \begin{eqnarray}
        \bm{E}_0f =& \begin{bmatrix}0 \\ -f\end{bmatrix} \\
        \bm{E}_{1/2}f =& \begin{bmatrix}f/2 \\ -f/2\end{bmatrix} \\
    \end{eqnarray}
    or the non-linear variant
    \begin{eqnarray}    
        \Tilde{\bm{E}}f =& \begin{bmatrix}\max(f,0) \\ -\min(f,0)\end{bmatrix}.
        \label{eq:etilde}
    \end{eqnarray}
    The only property that we require of $\bm{E}$ is that it should be right-inverse to $\bm{D}$, viz.\ 
    \begin{equation}
        \bm{D}\bm{E} = \bm{I}.
        \label{eq:DE_I}
    \end{equation}
    This property expresses the fact that lifting a function from $\mathcal{V}$ into $\mathcal{V}\times\mathcal{V}$ (the action of $\bm{E}$) followed by collapsing back into the space $\mathcal{V}$ of physical fluxes (the action of $\bm{D}$) should not change the function we started with. It is easily verifiable that the previously proposed definitions of $\bm{E}$ satisfy this property.
    In matrix notation, the operators $\bm{E}_1$, $\bm{E}_0$, and $\bm{E}_{1/2}$ are given by the general formula
    \begin{eqnarray}
        \bm{E}_z = \begin{bmatrix}zI \\ (z-1)I\end{bmatrix}.
    \end{eqnarray}
    It is not possible to express $\Tilde{\bm{E}}$ as a matrix, as it is non-linear.
    
    Let us now consider the operator product with the opposite ordering, $\bm{C} = \bm{E}\bm{D}$. This operator, $\bm{C}:\mathcal{V}\times\mathcal{V}\rightarrow\mathcal{V}\times\mathcal{V}$ is a projector, by virtue of Eq.~\eqref{eq:DE_I}:
    \begin{equation}
        \bm{C}^2 = \bm{E}\bm{D}\bm{E}\bm{D} = \bm{E}(\bm{D}\bm{E})\bm{D} = \bm{E}\bm{D} = \bm{C}.
        \label{eq:C2_C}
    \end{equation}
    By construction, the null space of $\bm{C}$ coincides with the null space of $\bm{D}$, which is the space of even (unphysical) vectors:
    \begin{equation}
        \bm{C}\begin{bmatrix}\eta \\ \eta\end{bmatrix} = 0.
    \end{equation}
    Therefore, the operator $\bm{C}$ can be regarded as a model for perfect cancellation. We have shown that the even vectors represent the solutions of the nonphysical Boltzmann equation in Eq.~\eqref{eq:modified_boltzmann}. This is supported by the explicit form of $\bm{C}$ which follows from the previous definitions for $\bm{E}$:
    \begin{eqnarray}
        \bm{C}_z \begin{bmatrix}f_+ \\ f_-\end{bmatrix} =& \bm{E}_z\bm{D}\begin{bmatrix}f_+ \\ f_-\end{bmatrix} =& \begin{bmatrix} z(f_+ - f_-) \\ (1-z)(f_+-f_-)\end{bmatrix} \\
        \Tilde{\bm{C}} \begin{bmatrix}f_+ \\ f_-\end{bmatrix} =& \Tilde{\bm{E}}\bm{D}\begin{bmatrix}f_+ \\ f_-\end{bmatrix} =& \begin{bmatrix} \max(f_+ - f_-, 0) \\ -\min(f_+-f_-,0)\end{bmatrix}.
        \label{eq:ctilde}
    \end{eqnarray}
    Note that only $\Tilde{\bm{C}}$ can guarantee that both vector components are non-negative, but it requires that both $\bm{C}$ and $\bm{E}$ be non-linear. In matrix notation, $\bm{C}_z$ reads
    \begin{equation}
        \bm{C}_z = \begin{bmatrix}zI & -zI \\ (z-1)I & -(z-1)I\end{bmatrix}.
        \label{eq:linear_cancellation}
    \end{equation}
    
    Going back to our eigenvalue equation, Eq.~\eqref{eq:cpld_mtrx} may be modified to include the application of cancellation on the fission source:
    \begin{equation}
        \bm{A}\bm{\zeta} = \frac{1}{k}\bm{C}\bm{F}\bm{\zeta}.
        \label{eq:cpld_mtrx_cancel}
    \end{equation}
    If $\bm{\zeta}$ is even, $\bm{F}\bm{\zeta}$ must also be even as $\comm{\bm{F}}{\bm{P}}=0$. Since $\bm{C}$ maps even vectors to $0$, the cancellation operator in Eq.~\eqref{eq:cpld_mtrx_cancel} causes the nonphysical eigenmodes corresponding to Eq.~\eqref{eq:modified_boltzmann} to vanish.
    
    At this point, it remains to be shown that cancellation does not perturb the physical eigenvalues of Eq.~\eqref{eq:cpld_mtrx}, that were initially associated with the odd eigenstates. To accomplish this, we start with Eq.~\eqref{eq:cpld_mtrx_cancel}, and apply $\bm{D}$ from the left on both sides, as this is the operator which performs the mapping $\mathcal{V}\times\mathcal{V}\rightarrow\mathcal{V}$:
    \begin{equation}
        \bm{D}\bm{A}\bm{\zeta} = \frac{1}{k}\bm{D}\bm{C}\bm{F}\bm{\zeta}.
    \end{equation}
    From Eq.~\eqref{eq:DE_I}, we may substitute $\bm{D}\bm{C}=\bm{D}$. Using Eqs.~\eqref{eq:zeta_deff}, \eqref{eq:A}, and \eqref{eq:phys_flux_from_weighted}, this then simplifies to
    \begin{equation}
        (\mathcal{A}_{11} - \mathcal{A}_{12})\varphi = \frac{1}{k}\mathcal{F}\varphi.
    \end{equation}
    Eq.~\eqref{eq:diode} indicates that $\Delta(x)-\Delta(-x) = x$, which, when combined with Eq.~\eqref{eq:A11_A12}, yields
    \begin{equation}
        \big[\hat{\bm{\Omega}}\cdot\nabla + \Sigma_t\big]\varphi = \mathcal{S}\varphi +  \frac{1}{k}\mathcal{F}\varphi.
    \end{equation}
    This is the physical Boltzmann transport equation. Therefore, cancellation leaves the physical eigenstates unchanged, so long as Eq.~\eqref{eq:DE_I} holds true, regardless of the form of $\bm{E}$.
    
    In summary, the introduction of cancellation conserves all the (physical and nonphysical) eigenstates of Eq.~\eqref{eq:cpld}; however, the nonphysical part of the spectrum is relegated to zero, while the physical eigenvalues are unperturbed and dominate the power iteration.
    
    \subsection{Deterministic proof of concept}
    In order to better understand the effects of cancellation, and how it positively affects the convergence of the power iteration, it is fruitful to visualize the spectrum of the associated equations. For this purpose, we have written a deterministic solver that computes the full spectrum of eigenvalues and eigenfunctions (using the LAPACK library \cite{lapack}) for a discretized version of the one-dimensional system introduced in Sec.~\ref{sec:pid}. A deterministic solver was necessary, since Monte Carlo can only be used to estimate the fundamental mode by power iteration. This solver uses a finite-difference method, with cross section values being taken at the midpoint of each spatial bin; both the positive and negative particle fluxes are explicitly considered.
    
\begin{figure}
        \includegraphics[width=\columnwidth]{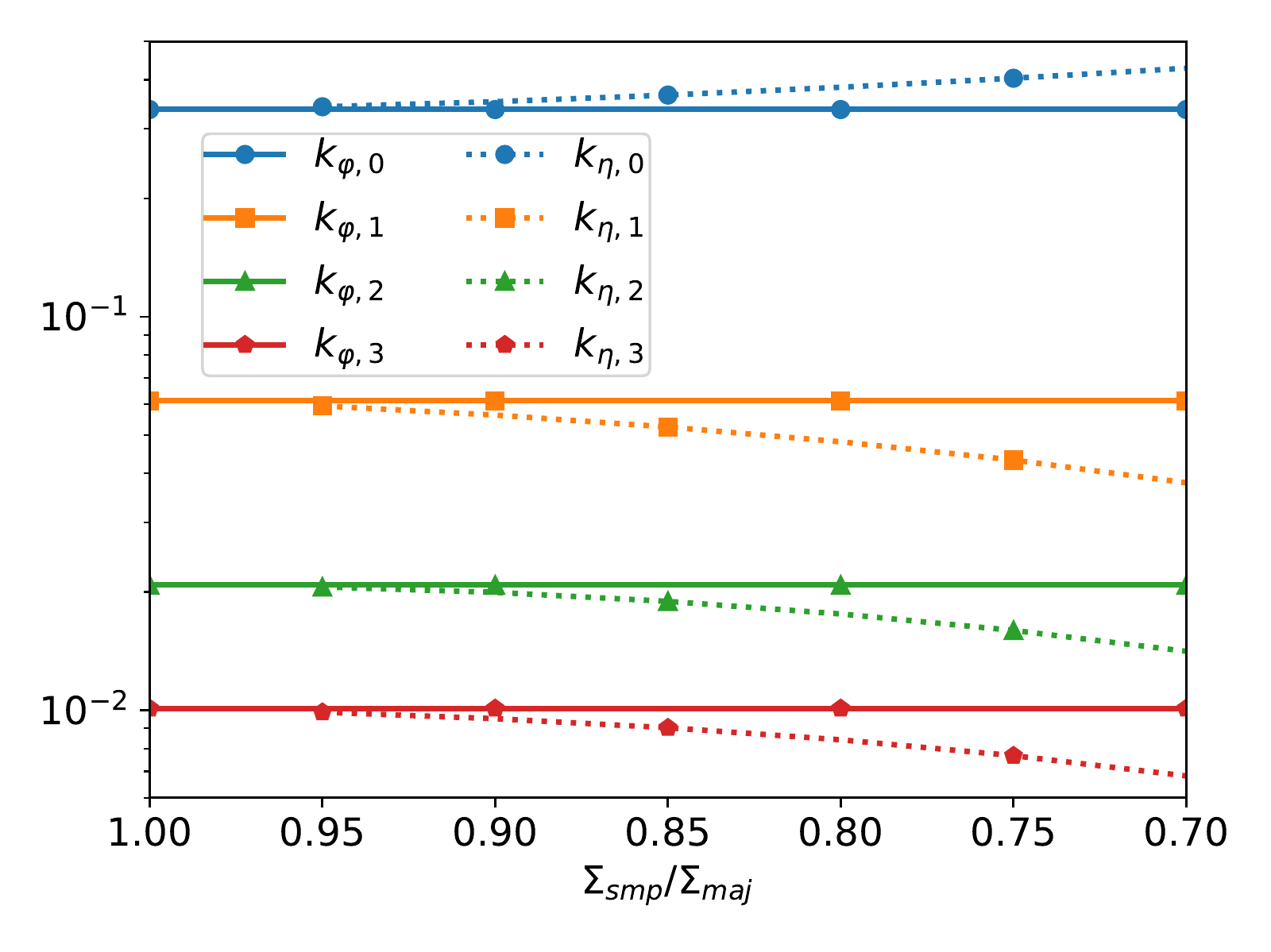}
        \caption{Behavior of the eigenvalue spectrum as a function of the ratio of the sampling cross section to the majorant cross section. Eigenvalues associated with the physical Boltzmann equation have a solid line, while those associated with the nonphysical Boltzmann equation have a dotted line.}
        \label{fig:k_spectrum}
    \end{figure}

    Before considering cancellation, it is helpful to observe the behaviour of the eigenvalue spectrum of Eq.~\eqref{eq:cpld_mtrx} as a function of $\Sigma_\text{smp}/\Sigma_\text{maj}$. This quantity roughly measures of the amount of negative weight that is produced by NWDT; the smaller the ratio is, the more negative weight is introduced. The spectrum is the union of the sets of eigenvalues $k_{\varphi,i}$ and $k_{\eta,i}$ and is presented in Figure~\ref{fig:k_spectrum}. Naturally, the spectrum associated with the physical Boltzmann equation does not change with the ratio of $\Sigma_\text{smp}/\Sigma_\text{maj}$. The eigenvalues of the nonphysical Boltzmann equation diverge from the physical values as this ratio decreases from unity. The dominant nonphysical eigenvalue $k_{\eta,0}$ continually increases above $k_{\varphi,0}$, while the other depicted eigenvalues decrease from their physical counterparts. In Eq.~\eqref{eq:kn_gt_keff}, we heuristically demonstrated that $k_{\eta,0}\ge k_\text{eff}$, and this is indeed the case in Figure~\ref{fig:k_spectrum}. Our heuristics however, are not able to make any remarks as to the behavior of the higher eigenvalues, and while for this system $k_{\eta,i}$ for $i>0$ always appears to decrease when $\Sigma_{smp}/\Sigma_{maj}$ decreases, we have observed other systems which do not exhibit this behavior.
    
    When the cancellation operator $\bm{C_1}$ is added in accordance with Eqs.~\eqref{eq:linear_cancellation} and \eqref{eq:cpld_mtrx_cancel}, we find that cancellation indeed suppresses the even eigenstates, leaving only the odd ones associated with the physical solutions, so long as $\Sigma_\text{smp}/\Sigma_\text{maj} > 0$. This is the case for a cancellation operator which is ``perfect'', in that it is able to conduct cancellation in a manner which always completely neutralizes the negative particle population. While this is of course desired, and possible to implement in a deterministic solver, we will later show in Sec.~\ref{sec:1d_cancel} that 100\% cancellation efficiency is not necessarily achievable in a Monte Carlo approach. To mimic this fact, we introduce an imperfect cancellation operator $\bm{C}^{(\alpha)}$:
    \begin{eqnarray}
        \bm{C}^{(\alpha)} = \alpha\bm{C}_1 + (1-\alpha)\bm{I}.
    \end{eqnarray}
    Here $\alpha$ represents the cancellation efficiency: when $\alpha=1$ there is perfect cancellation, and when $\alpha=0$ there is no cancellation. This is a highly idealized approach to model imperfect cancellation, and is not necessarily a faithful model of the effect of cancellation in a Monte Carlo setting.
    
    \begin{figure}
        \includegraphics[width=\columnwidth]{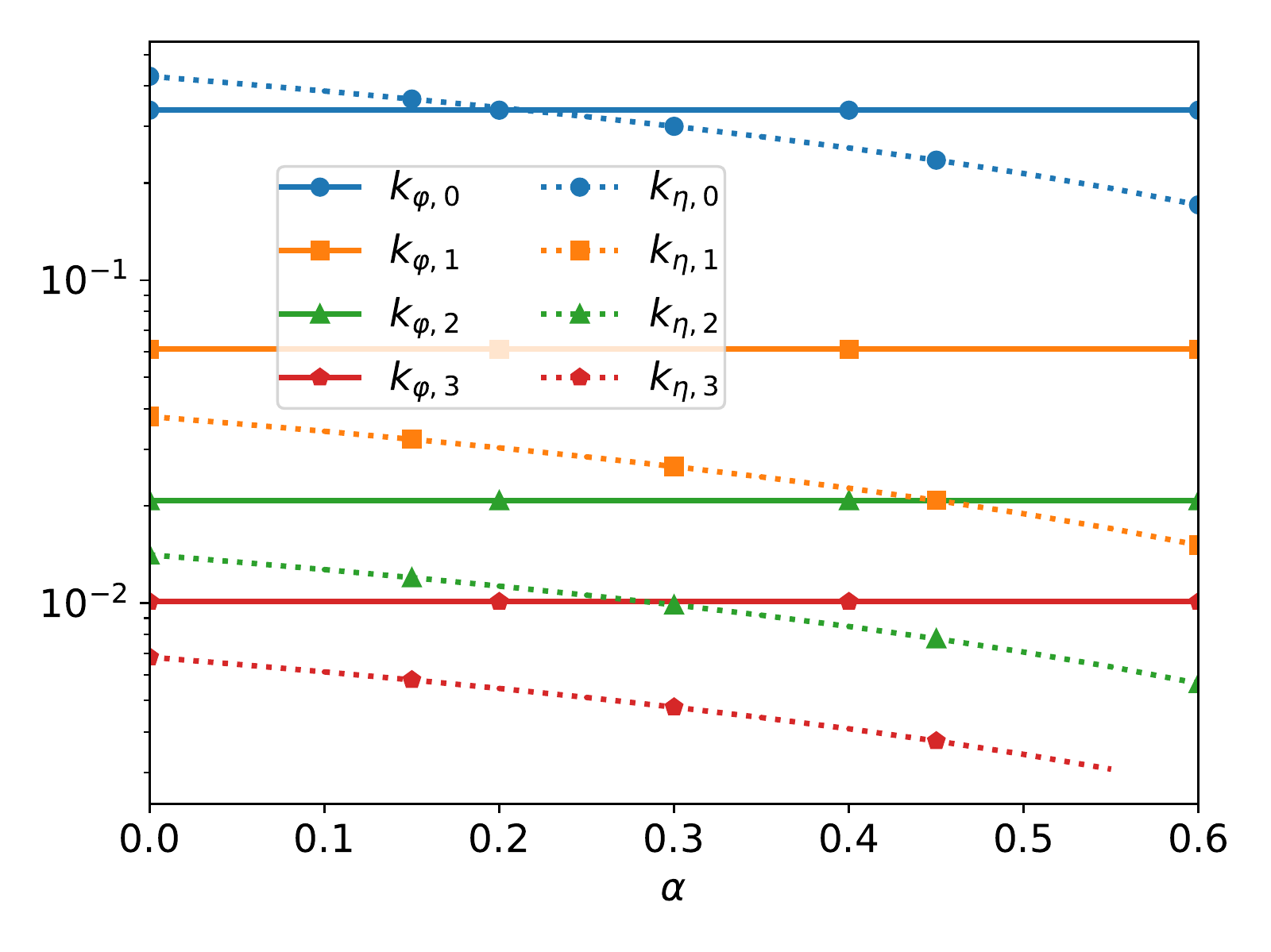}
        \caption{Behaviour of the eigenvalue spectrum as a function of cancellation efficiency $\alpha$, for the case of $\Sigma_\text{smp}/\Sigma_\text{maj}=0.7$. Eigenvalues associated with the physical Boltzmann equation have a solid line, while those associated with the nonphysical Boltzmann equation have a dotted line.}
        \label{fig:alpha}
    \end{figure}
    
    Figure~\ref{fig:alpha} presents the effects of cancellation with varying levels of efficiency, for the case of $\Sigma_\text{smp}/\Sigma_\text{maj}=0.7$. For values of $\alpha$ between 1 and approximately 0.21, the cancellation operator remains efficient enough that the physical eigenstate is the dominant one. A critical point is reached near $\alpha\approx0.21$, where the eigenvalues of the physical and nonphysical systems are equal. Further reducing the efficiency of cancellation leads to the nonphysical eigenstate being the dominant one. This example indicates that for our system, there is a \emph{minimum amount of weight cancellation} required in order for power iteration to converge on the fundamental physical eigenstate. This is likely true for all systems which can be described by coupled transport equations for positive and negative particles. 
    
    Even in settings where enough cancellation is present to make the $k_{\varphi,0}$ mode dominant, cancellation might still be unable to reduce $k_{\eta,0}$ to a level that allows for rapid statistical convergence towards the fundamental mode. To clarify this, we introduce the dominance ratio, which is defined as the ratio of the second-largest eigenvalue to the largest one. The dominance ratio is equal to $k_{\eta,0}/k_{\varphi,0}$ for values of $\alpha$ where $k_{\varphi,0} > k_{\eta,0} > k_{\varphi,1}$. If instead $\alpha$ is such that $k_{\eta,0} < k_{\varphi,1}$, then the dominance ratio is $k_{\varphi,1}/k_{\varphi,0}$, which is the dominance ratio of the physical system. The dominance ratio provides an indication of the rate of convergence of the power iteration. When it is very close to unity, more iterations are required to converge on the dominant eigenvalue. Should $\alpha$ be sufficiently large to ensure $k_{\eta,0} < k_{\varphi,1}$, the convergence rate of the problem will no longer be bound by the efficiency of cancellation, but by the physical properties of the system being examined.
    
    This model for cancellation in a deterministic context provides valuable insight as to the behavior of this coupled system of positive and negative particles, and to the possible behavior of implementing cancellation of particle weights in a Monte Carlo context. It is for this reason that we do not go beyond the provided surface-level analysis of the effects of cancellation efficiency or choice of $\Sigma_\text{smp}$ on the convergence of the deterministic model. We will now continue by discussing techniques of weight cancellation in Monte Carlo simulations.
    
    \section{An Exact Regional Cancellation Scheme for 1D Problems}
    \label{sec:1d_cancel}
    
    As mentioned in the introduction, several cancellation strategies have been proposed in the past; one of these, devised by Booth and Gubernatis, is exact in 1D geometries \cite{booth2010exact}. This method works by partitioning all fissile domains of the problem domain into regions. Between each generation, weight cancellation amongst all new fission particles born in the region occurs. An important feature of this algorithm is that it has linear computational complexity with the number of particles partaking in cancellation, making it a good candidate for inclusion in general-purpose Monte Carlo transport codes. We have therefore chosen to focus on this algorithm. We will now provide a brief overview of how Booth and Gubernatis' method works, as it is essential for understanding our proposed 3D algorithm, which will be developed  in Sec.~\ref{sec:3d_cancel}.
    
   Booth and Gubernatis make use of a quantity referred to as the fission density \cite{booth2010exact}. In an effort to reproduce and expand upon their work, we have chosen the following definition for the expected fission density: considering a particle starting a flight at position $x_0$ and traveling in direction $\mu=\pm 1$, its expected fission density $f(x|x_0,\mu)$ is the expected value of the number of fission events per unit length around $x$. It can be written as the product of the probability density of flying from $x_0$ to $x$ and having a collision at position $x$, and the probability of that collision being in the fission reaction channel. The exact form of $f(x|x_0,\mu)$ depends on the transport methodology being employed. For NWDT, this formula can be deduced by examination of Alg.~\ref{alg:ndwt}. The probability density of flying from $x_0$ to $x$ and having a real collision is
    \begin{equation}
        P_c(x|x_0,\mu) = \begin{cases} q \frac{\Sigma_t(x)}{q\Sigma_\text{smp}}\Sigma_\text{smp}e^{-\Sigma_\text{smp}\abs{x-x_0}} & \frac{x-x_0}{\abs{x - x_0}} = \mu \\\\ 0 & \frac{x-x_0}{\abs{x - x_0}} \not= \mu \end{cases}.
    \end{equation}
    The $\Sigma_\text{smp}e^{-\Sigma_\text{smp}\abs{x-x_0}}$ portion is the probability density of flying from $x_0$ to $x$, and having either a real or virtual collision. The factor $q$ is the probability of the collision being real, while $\Sigma_t(x)/(q\Sigma_\text{smp})$ is the weight correction factor for real collisions. This must of course be combined with the fission probability
    \begin{equation}
        P_f(x) = \frac{\Sigma_f(x)}{\Sigma_t(x)}.
    \end{equation}
    As $f(x|x_0,\mu) = P_f(x)P_c(x|x_0,\mu)$, for the case of NWDT we may write
    \begin{equation}
        f(x|x_0,\mu) = \begin{cases}\Sigma_f(x) e^{-\Sigma_\text{smp}(x_0)\abs{x - x_0}} & \frac{x-x_0}{\abs{x - x_0}} = \mu \\\\ 0 & \frac{x-x_0}{\abs{x - x_0}} \not= \mu \end{cases},
        \label{eq:1d_f}
    \end{equation}
    noting that the fission density is zero for all positions which cannot be reached by the particle during the considered flight.
    
    \begin{figure}
        \includegraphics[width=\columnwidth]{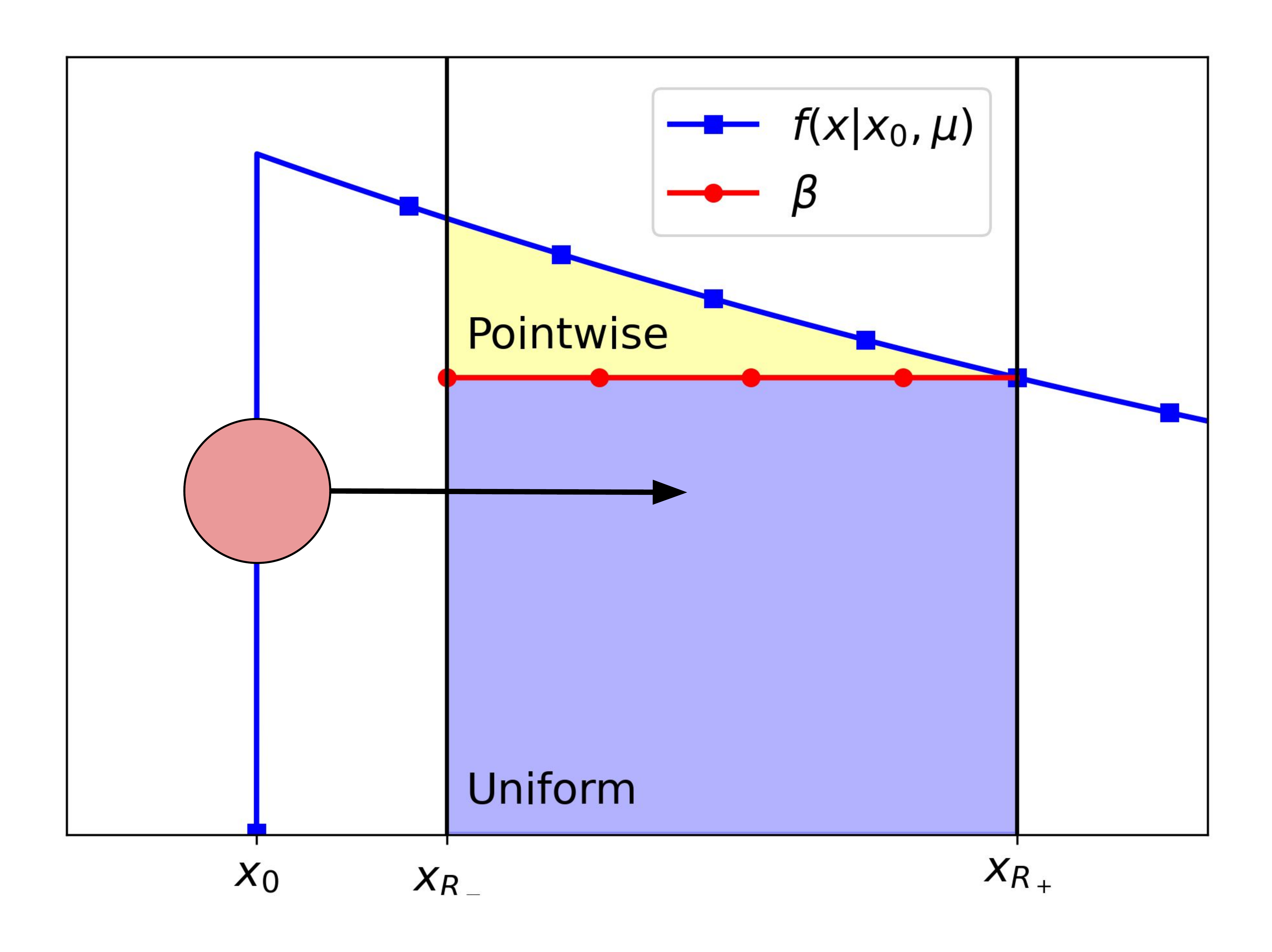}
        \caption{Presented is a neutron (the colored circle), beginning a flight at $x_0$, and flying in the $\mu=1$ direction (direction of flight is indicated by the arrow). For this flight, the expected fission density, $f(x|x_0,\mu)$ has been plotted as a blue line. The limits of the region are denoted at $x_{R_-}$ and $x_{R_+}$. The value of $\beta$ displayed here is the minimum value of $f(x|x_0,\mu)$ within the region. Note that $f(x|x_0,\mu) = 0$ for positions that cannot be reached by the particle.}
        \label{fig:fd}
    \end{figure}
    
    With the expected fission density for a particle having been defined, let us consider a fission particle $p$ belonging to the fission source and ready to start a random walk, at position $x_p$, located in an arbitrary region $R$ (with bounds $x_{R_-}$ and $x_{R_+}$). The parent of this particle (which was a member of the previous fission generation) was previously at position $x_0$ before flying to $x_p$, and producing the current particle of interest. Figure~\ref{fig:fd} illustrates the expected fission density of the parent particle, as well as the bounds of region $R$, and $x_0$. The depicted value of $\beta$ in the plot is the minimum value of the fission density within the region, for the flight of the parent particle:
    \begin{equation}
        \beta = \min_{x\in R} f(x|x_0,\mu).
    \end{equation}
    The idea of Booth and Gubernatis' method is to estimate the fission density associated with the flight as the sum of a uniform component over the region $R$ and a pointwise component concentrated at the actual fission site. For a fission particle born in $R$, which was induced by a particle beginning a flight at $x_0$ and traveling in the direction $\mu$, the bottom shaded portion in Fig.~\ref{fig:fd} represents the fraction of the fission density which is uniformly distributed in $R$, while the top portion represents the non-uniform fraction of the fission density, which depends on the position $x_p$ where the parent lands and induces a fission. This indicates that, for the particle $p$, the fraction of the fission density concentrated at $x_p$ is $(f(x_p|x_0,\mu)-\beta)/f(x_p|x_0,\mu)$, and the fraction uniformly distributed through region $R$ is $\beta/f(x_p|x_0,\mu)$. As such, we can set the weight of $p$ to be
    \begin{equation}
        w_p = w \frac{f(x_p|x_0,\mu) - \beta}{f(x_p|x_0,\mu)}
        \label{eq:wp}
    \end{equation}
    and at the same time create a uniform weight portion $w_u$ of the particle, representing the portion of the particle which is evenly distributed throughout $R$, namely
    \begin{equation}
        w_u = w\frac{\beta}{f(x_p|x_0,\mu)}.
        \label{eq:wu}
    \end{equation}
    It is important to note that $w_p + w_u = w$: the net weight in the system has not been modified, we have simply displaced a portion of $w$, distributing it uniformly through region $R$.
    
    Dividing particles into a pointwise and uniform portion does not in and of itself accomplish any weight cancellation. When there are many fission particles within the same region, however, all of their uniform weight portions may be combined into a single weight which represents the uniformly distributed portion of the fission source for the region. This quantity shall be denoted as $U_R$, and is the sum of all uniform weight portions for all particles born within the region, each coming with a sign. Negative particles will yield negative contributions to $U_R$, while positive particles will yield positive contributions, leading to a cancellation. Once all particles have contributed their uniform portion to $U_R$, $n$ new particles are sampled uniformly within the region, where
    \begin{equation}
        n = \left\lceil \abs{U_R} \right\rceil.
    \end{equation}
    Each of these uniformly sampled particles within $R$ then has a weight of $U_R/n$.
    
    These newly sampled, uniformly distributed particles belong to the same generation as particle $p$, and behave exactly like traditionally generated particles from this point on. Under this scheme, a particle which was initially positive will produce a positive uniform portion, and keep a positive weight; conversely, negative particles will remain negative and produce negative uniform portions. In the case of $x_0$ being in the same region as $x_p$, the uniform weight portion $w_u$ must be zero, as the minimum of the expected fission density ``behind'' the starting point of the flight is zero. Therefore, particles with $x_0$ and $x_p$ in the same region effectively do not partake in the cancellation process. These two properties are ensured by $\beta$ being the minimum value of $f$ in the region. However, it is possible to relax the requirement that $\beta$ should be the minimum of the expected fission density of the region. Booth and Gubernatis show in their original work that any value of $\beta$ may be used, while still producing an unbiased result \cite{booth2010exact}. In the event of $\beta > f(x_p|x_0,\mu)$, the sign of the particle's pointwise weight portion will change.
    
    With this regional method, the efficiency of cancellation can never be 100\%. Only the uniform weight portions contribute to cancellation. For negative weighted fission particles, the pointwise weight portions do not cancel, always leaving some residual negative weight%
    %(unless the particle is located exactly at the position where $f(x_p|x_0,\mu) = \beta$)
    . Given this information, one might think that using a value of $\beta$ which is larger than the minimum fission density might be beneficial, as it would increase the magnitude of the weight which goes into the uniform portion. This is not necessarily the case, however, as it could lead to the pointwise portions of positive particles becoming negative, via Eq.~\eqref{eq:wp}, possibly defeating the purpose of cancellation. Sadly, this method of cancellation can not be modeled as a linear operator, and therefore we are unable to put it in the context of of the framework which was put forth in Section~\ref{sec:theory}. This makes regional cancellation conceptually similar to (though not the same as) the $\tilde{\bm{C}}$ operator mentioned in Eq.~\eqref{eq:ctilde}. It is inconsequential if our cancellation is linear or not; so long as it corresponds to a choice of $\bm{E}$ that preserves Eq.~\eqref{eq:DE_I}, the method will be exact and unbiased.
    
    \subsection{Results of the 1D methodology}
    
\begin{figure}
        \includegraphics[width=\columnwidth]{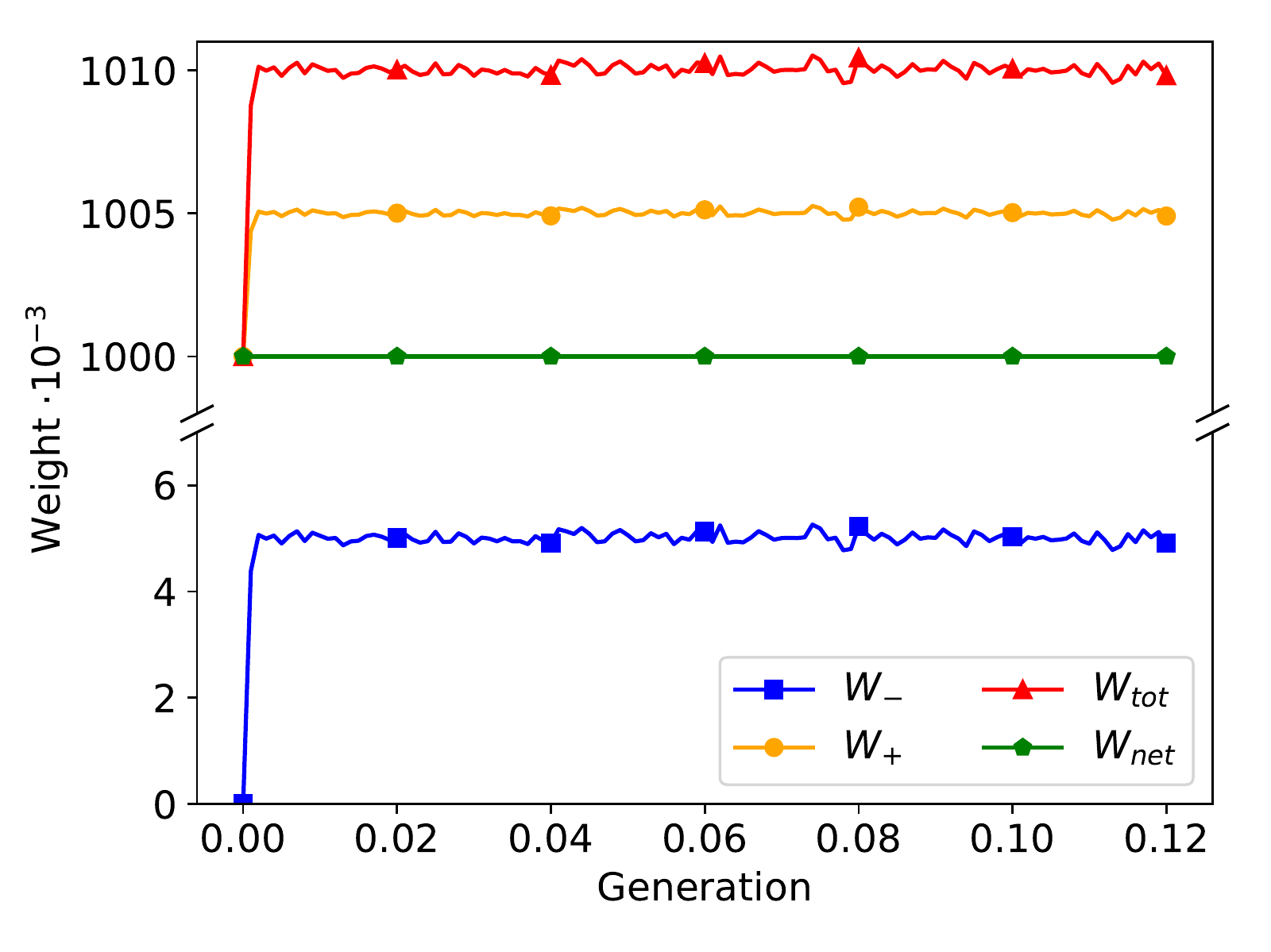}
        \caption{The positive weight ($W_+$), negative weight ($W_-$), net weight ($W_\text{net}$), and total weight ($W_\text{tot}$) in the 1D rod system, when using 30 cancellation regions and $10^6$ initial particles.}
        \label{fig:cancel_weights}
    \end{figure}

\begin{figure}
        \includegraphics[width=\columnwidth]{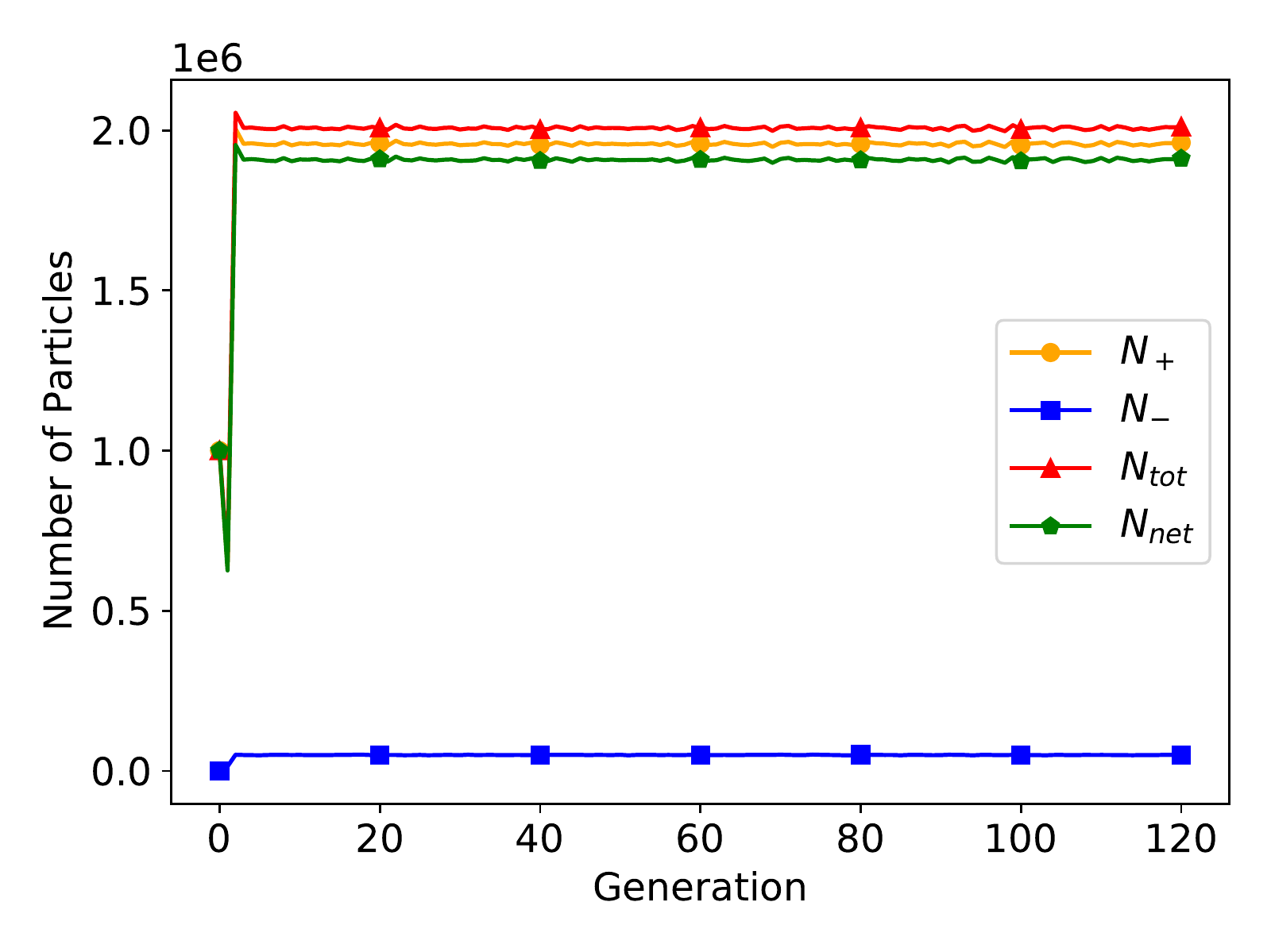}
        \caption{The number of positive ($N_+$), negative ($N_-$), net ($N_\text{net}$), and total ($N_\text{tot}$) particles in the 1D rod system, when using 30 cancellation regions and $10^6$ initial particles.}
        \label{fig:cancel_nums}
    \end{figure}

    We added the exact 1D cancellation method of Booth and Gubernatis to the power iteration problem outlined in Sec.~\ref{sec:pid}, performing the cancellation between generations, and before the normalization of the system weight occurs. Thirty evenly spaced cancellation regions were used to partition the rod, and $\beta$ was always taken to be the minimum of the expected fission density within the region. With cancellation, the simulation was stable, and was able to complete without issue. An eigenvalue of $k_\text{eff} = 0.33577 \pm 0.00005$ was obtained after 120 generations (20 inactive generations). This is in excellent agreement with the value obtained for the system using delta tracking, which was $k_\text{eff} = 0.33573 \pm 0.00005$ after the same number of generations. The weights $W_+$, $W_-$, $W_\text{net}$, and $W_\text{tot}$ are shown in Figure~\ref{fig:cancel_weights}.
    We see that the negative weight quickly rises to an equilibrium level, near 5000, and fluctuates then about that value for the duration of the simulation. The positive weight must naturally increase by the same amount to keep the net weight of the system constant. This leads to an increase in the total transported weight of approximately $10^4$ (the net weight still being $10^6$). The evolution of the number of positive and negative particles is presented in Figure~\ref{fig:cancel_nums}. The number of particles in memory stabilizes near almost $2\cdot 10^6$, twice as many particles as were initially used.%As there is comparatively very little extra weight in the system ($W_\text{tot} \approx W_\text{net}$), this indicates that the average weight of a positive particle between generations is approximately $0.5$. %This could be changed by applying roulette or particle combing after the cancellation procedure, before the next generation has started, but we chose not to do this, as particles already undergo roulette during the random walk.
    
\begin{figure}
        \includegraphics[width=\columnwidth]{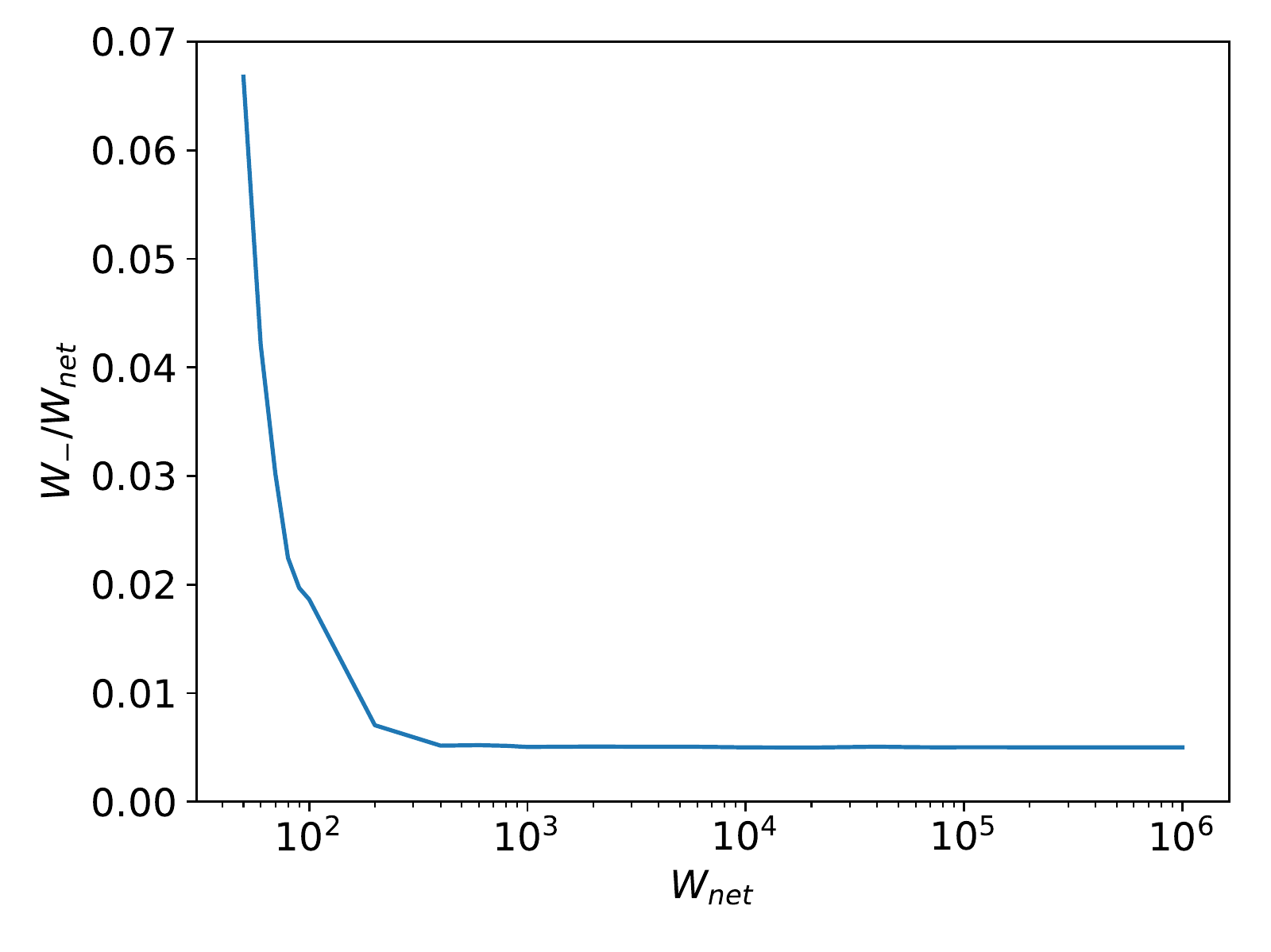}
        \caption{Negative weight fraction ($W_-/W_\text{net}$) as a function of net weight ($W_\text{net}$) in the 1D rod system, when using cancellation with 30 cancellation regions.}
        \label{fig:wnet_wnet}
    \end{figure}

    The cancellation algorithm itself has linear computational complexity with respect to the number of fission particles at any given generation. However, it is less obvious whether the total number of fission particles present once the simulation has settled to equilibrium is itself a linear function of the net weight of the system. It is evident that the total amount of weight transported must increase until equilibrium is reached, but is the increase linear in the net weight of the system? To examine this, we performed several runs with different initial values of $W_\text{net}$ and we looked at the average value of $W_-/W_\text{net}$, as a function of $W_\text{net}$. The total weight can be deduced by looking at only the behavior of the negative weight, as the increase in the positive weight will mirror the negative weight. The result of this study is presented in Fig.~\ref{fig:wnet_wnet}.
    When too little net weight is injected in the system, a higher percentage of negative weight will be present in the system at equilibrium. There appears to be a critical point (in this case, near $W_\text{net}=400$) above which the fraction of negative weight no longer decreases by adding more net weight. This might at first seem counter-intuitive, but positive particles are always being converted to negative particles during the random walk (because of the NWDT algorithm), and the cancellation process is not 100\% efficient. Above this critical point, no matter how much positive weight is added to the system at the beginning of the simulation, the equilibrium amount of negative weight will on average be a set fraction of the net weight. In other words, the total equilibrium weight is a linear function of the starting weight only for sufficiently large values of the latter.
    
\begin{figure}
        \includegraphics[width=\columnwidth]{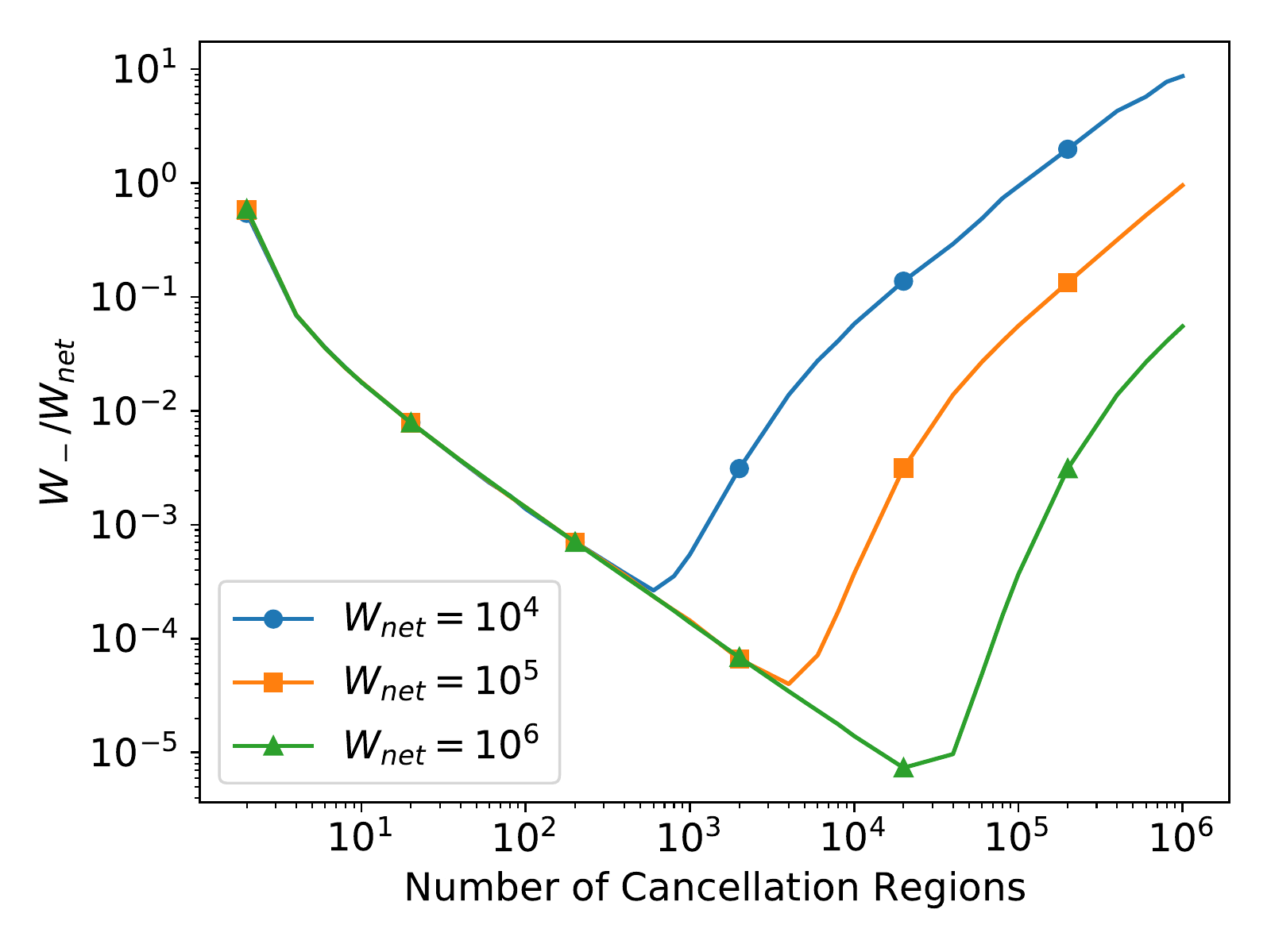}
        \caption{Negative weight fraction ($W_-/W_\text{net}$) as a function of the number of cancellation regions, for different net weights.}
        \label{fig:bins}
    \end{figure}

    Another consideration is the effect of the number of cancellation regions on the amount of negative weight. When there are fewer regions, each one must become larger; this makes cancellation less efficient, as $\beta$ will decrease if one is taking $\beta$ to be the minimum of $f$ over the region. Conversely, making regions too small will result in too few particles which can partake in cancellation, also reducing the efficiency. To examine this, we have plotted the negative weight fraction for the number of cancellation regions in Figure~\ref{fig:bins}.
    Indeed, from the figure, there is a range for which the fraction of negative weight decreases (i.e.\ the efficiency of cancellation increases) by adding more regions. The number of regions where the minimum occurs depends on the net weight of the system. Here, this occurs near \num{6e2} regions for $W_\text{net}=\num{e4}$, \num{4e3} for $W_\text{net}=\num{e5}$ and \num{2e4} for $W_\text{net}=\num{e6}$, suggesting that the optimal number of regions does not quite scale linearly with the net initial weight. The three tested net weights all had very similar behavior, and the efficiency improved at the same rate when adding cancellation regions, until the minimum value was reached. There is also a large range over which the negative weight fraction is less than 1\% of the net weight. For all net weights, this range starts at approximately 20 regions, and goes up to \num{2e3} regions for $W_\text{net}=\num{e4}$, or \num{3e5} regions for $W_\text{net}=\num{e6}$.
    
    This section has outlined the methodology behind Booth and Gubernatis' regional cancellation. The potential of the method was successfully demonstrated, having canceled enough negative weight to allow for the proper convergence of the power iteration algorithm when using NWDT. In light of this success, we move on to extending the method to work with more realistic problems in higher spatial dimensions.
    
    \section{Exact 3D, Multi-Group, Regional Cancellation Scheme}
    \label{sec:3d_cancel}
    
    Unfortunately, it is not possible to directly implement the 1D regional cancellation scheme of Booth and Gubernatis in higher spatial dimensions. One-dimensional systems have the special property that lines coincide with volumes, ensuring the lines of flight of all particles traversing the region overlap, as well as their uniform weight portions. This fact allows us to add the uniform weight portions of all particles in the region, effectively leading to a cancellation of negative and positive weight. For higher dimensions, this is not the case. The uniform portion of the particle weights may only be distributed along their rays of flight within the cancellation region, but the flight rays will never overlap completely. It is then no longer possible to combine the uniform weight portions as before by simply taking their sum.
    
    The problem with the method of calculating the fission density in Eq.~\eqref{eq:1d_f} is that it is only valid along the ray of flight of the parent particle, after the flight direction has been sampled. To extend the regional cancellation to higher dimensions, it is required to consider the probability of the parent particle scattering into the solid angle which is subtended by the cancellation region. This is done by examining $\hat{\bm{\Omega}}'$, the direction of the parent particle at position $\bm{r}_0$ before the direction is changed by the scattering kernel. After the parent has been modified by the process of scattering, it then has a direction $\hat{\bm{\Omega}}$, which must intersect the cancellation region (given that we are attempting to perform cancellation for a particle in the region which was induced by the parent). The scattering cosine for the interaction which the parent underwent at $\bm{r}_0$ is then $\mu = \hat{\bm{\Omega}}'\cdot\hat{\bm{\Omega}}$. For neutron transport, the azimuthal direction of a scatter is almost always isotropic, and only the scattering cosine is anisotropic (should there be any anisotropy) \cite{bell_glasstone}. The probability density of scattering in direction $\hat{\bm{\Omega}}$ from direction $\hat{\bm{\Omega}}'$ is therefore
    \begin{equation}
        P({\hat{\bm{\Omega}}}|{\hat{\bm{\Omega}}}') = \frac{P(\mu)}{2\pi},
    \end{equation}
    with $\int_{-1}^{1} P(\mu) d\mu = 1$. Modifying Eq.~\eqref{eq:1d_f} to use the flight kernel for 3D, and applying this factor to consider the scattering angle, we arrive at the expected fission density
    \begin{eqnarray}
        f(\bm{r}|\bm{r}_0,\hat{\bm{\Omega}}') = \frac{P(\mu)\Sigma_f(\bm{r})}{2\pi\abs{\bm{r} - \bm{r}_0}^2} e^{-\Sigma_\text{smp}(\bm{r}_0)\abs{\bm{r} - \bm{r}_0}},
        \label{eq:3d_f}
    \end{eqnarray}
    where
    \begin{eqnarray}
        \mu = \frac{\bm{r} - \bm{r}_0}{\abs{\bm{r} - \bm{r}_0}} \cdot \hat{\bm{\Omega}}'.
    \end{eqnarray}
    Taking a closer look at Eq.~\eqref{eq:3d_f}, one may recognize it as being what Lux and Koblinger refer to as the next-event estimator, for the fission rate at $\bm{r}$ given a collision at $\bm{r}_0$ \cite[][Sec.~6.IV.A]{lux}. Note that Eq.~\eqref{eq:3d_f} represents the expected fission density over the scattering and the free flight following it; this should be contrasted with Eq.~\eqref{eq:1d_f}, which represents the expected fission density over the following free flight \emph{only}.
    
    \begin{figure}
        \includegraphics[width=\columnwidth]{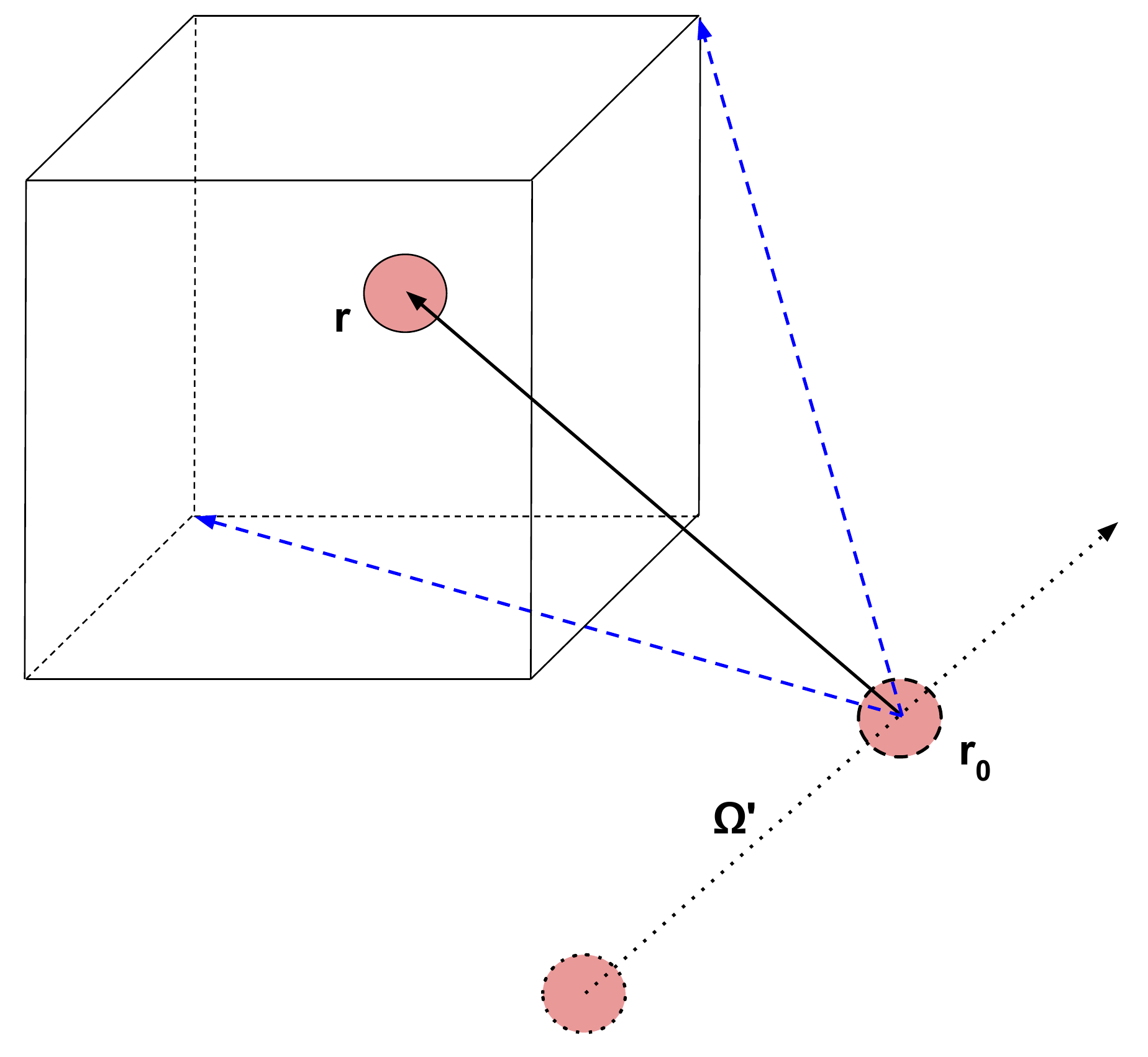}
        \caption{Depiction of the 3D regional cancellation process. To find the uniform weight fraction of the fission particle at $\bm{r}$, $\beta$ must first be determined. This is done by calculating the expected fission density at all eight corners of the cancellation region (though only two are depicted by the dashed arrows) from the parent particles previous position and direction ($\bm{r}_0$ and $\hat{\bm{\Omega}}'$), and taking the minimum value.}
        \label{fig:3d_cancel}
    \end{figure}
    
    While there is only one natural shape for a cancellation region in 1D, there are an infinite number of possible shapes in 3D which one could use. In this work, we will consider right rectangular prisms for our cancellation region. This is due to the fact that it is very simple to sample positions within a right rectangular prism uniformly, which is required in the cancellation processes. We will choose $\beta$ to be the minimum value of the expected fission density within the cancellation region. If $\Sigma_f(\bm{r})$ is spatially varying within the cancellation region and scattering is anisotropic, it is difficult to determine the true minimum; when $\Sigma_f(\bm{r})$ is homogeneous within the region and scattering is isotropic, however, one may determine the true minimum of the expected fission density for the flight from $\bm{r}_0$ to $\bm{r}$ by evaluating $f(\bm{r}|\bm{r}_0,\hat{\bm{\Omega}}')$ at all eight corners of the prism, as the minimum must occur at one of those eight points. Figure~\ref{fig:3d_cancel} provides a depiction of the cancellation process. Using the minimum as the value of $\beta$ for the flight, it is possible to calculate a pointwise weight portion and a uniform weight portion of the fission particle, as was done before in Eq.~\eqref{eq:wp} and Eq.~\eqref{eq:wu}. Once the uniform portions of all fission particles in the region have been collected, one may take the net uniform weight of the region, and sample new particles to add to the fission bank uniformly within the prism.
     
    Until now, for simplicity, we have only considered single-speed transport, with all cross sections being energy-independent. With no energy dependence, one does not need to consider how to sample the energy of the uniform particles. In all realistic  applications in reactor physics, however, all cross sections are continuous functions of the neutron energy, and the emission spectrum for the energy of fission neutrons depends on the collided nuclide and on the incoming energy of the neutron which induced the fission \cite{bell_glasstone}. Under this assumption, it would be impossible to collect uniform portions of all the fission particles together. Without this action, no cancellation occurs. This fact poses a difficulty in applying weight cancellation to continuous energy systems.
    
    An approximation which is often used in the reactor physics community is the multi-group treatment of the energy variable in the phase space, by using cross sections which are piece-wise constant in energy. These energy intervals in which the cross section is constant are referred to as groups. We denote the $g$-th energy group as $E_g$, with $g\in\{1,2, .., N_g\}$. By convention, the group corresponding with the highest energy is $E_1$, while the group with the lowest energy is $E_{N_g}$ \cite{bell_glasstone}. The multi-group $k$-eigenvalue Boltzmann equation reads
    \begin{widetext}
    \begin{eqnarray}
        \hat{\bm{\Omega}}\cdot\nabla\varphi(\bm{r},\hat{\bm{\Omega}},E_g) + \Sigma_t(\bm{r},E_g)\varphi(\bm{r},\hat{\bm{\Omega}},E_g) = \sum_{g' = 1}^{N_g}\int_{4\pi}\Sigma_s(\bm{r}, \hat{\bm{\Omega}}'\rightarrow\hat{\bm{\Omega}}, E_{g'}\rightarrow E_g)\varphi(\bm{r},\hat{\bm{\Omega}}',E_{g'})\dd\hat{\bm{\Omega}}' +\nonumber\\ \frac{1}{4\pi k}\sum_{g'=1}^{N_g}\chi(E_{g'}\rightarrow E_g)\nu(\bm{r},E_{g'})\Sigma_f(\bm{r},E_{g'})\int_{4\pi} \varphi(\bm{r},\hat{\bm{\Omega}}',E_{g'})\dd\hat{\bm{\Omega}}'
        \label{eq:mg_boltzmann}
    \end{eqnarray}
    \end{widetext}
    The structure of Eq.~\eqref{eq:mg_boltzmann} is formally that of a system of particles with $N_g$ species, coupled with each other by means of scattering or fission.
    
    Here, the fission spectrum $\chi(E_{g'}\rightarrow E_g)$ is written to show a dependence of the fission particle energy on the parent particle's incident energy. This dependence on the incoming particle energy is relatively weak and is often ignored in the multi-group approximation \cite{bell_glasstone}. In this case, we may simply write $\chi(E_g)$, moving it outside the sum in Eq.~\eqref{eq:mg_boltzmann}. Under this approximation, we may collect the uniform portions of all fission neutrons in the region together; regardless of the energy of their parent, their energy spectra are all the same. However, one must change the definition of the expected fission density to take energy into account. Equation~\eqref{eq:3d_f} is then modified to include the energy group $E_g$ of the particle that induces fission:
    \begin{eqnarray}
        f(\bm{r}|\bm{r}_0,\hat{\bm{\Omega}}',E_g) = \frac{P(\mu,E_g)\Sigma_f(\bm{r},E_g)}{2\pi\abs{\bm{r} - \bm{r}_0}^2} \nonumber \\ \times e^{-\Sigma_\text{smp}(\bm{r}_0,E_g)\abs{\bm{r} - \bm{r}_0}}.
        \label{eq:3d_f_mg}
    \end{eqnarray}
    
    A similar problem can occur for the direction of the fission particles as well. In continuous energy transport, both the direction and energy of the fission particles may be a function of the energy and direction of the incoming neutron. While the nuclear data representations allow for this,  it is a very marginal occurrence\footnote{One exception to this is the evaluation for \isotope[232]{Th} which has anisotropic distributions for prompt fission neutrons in the laboratory frame, in both ENDF/B-VIII.0 and JEFF-3.3.} in major nuclear-data evaluation libraries \cite{endf8,jeff33}. Therefore, we do not need to make any special considerations for the angular distribution of fission particles in order to apply regional cancellation to multi-group problems.
    
    We have seen that weight cancellation takes place at fission; however, one may wonder what is special about fission, and whether one could perform cancellation at scattering events, instead. At fission, the energy and angular distributions of the secondary fission particles are independent of the properties of the particle that induces the fission event. Fission represents the natural cancellation event because the three-dimensional distribution of the expected fission density (Eq.~\eqref{eq:3d_f_mg}) encodes all the six-dimensional distribution of the fission emission density in phase space.
    
    % You have not demonstrated anywhere that this method is exact. I think it is,
    % but with so much emphasis in the abstract on the fact that it is exact, you
    % really need to call attention to why that is explicitly.

    \section{Implementation and Results}
    \label{sec:results}
    
    \begin{figure*}
        \includegraphics[width=\textwidth]{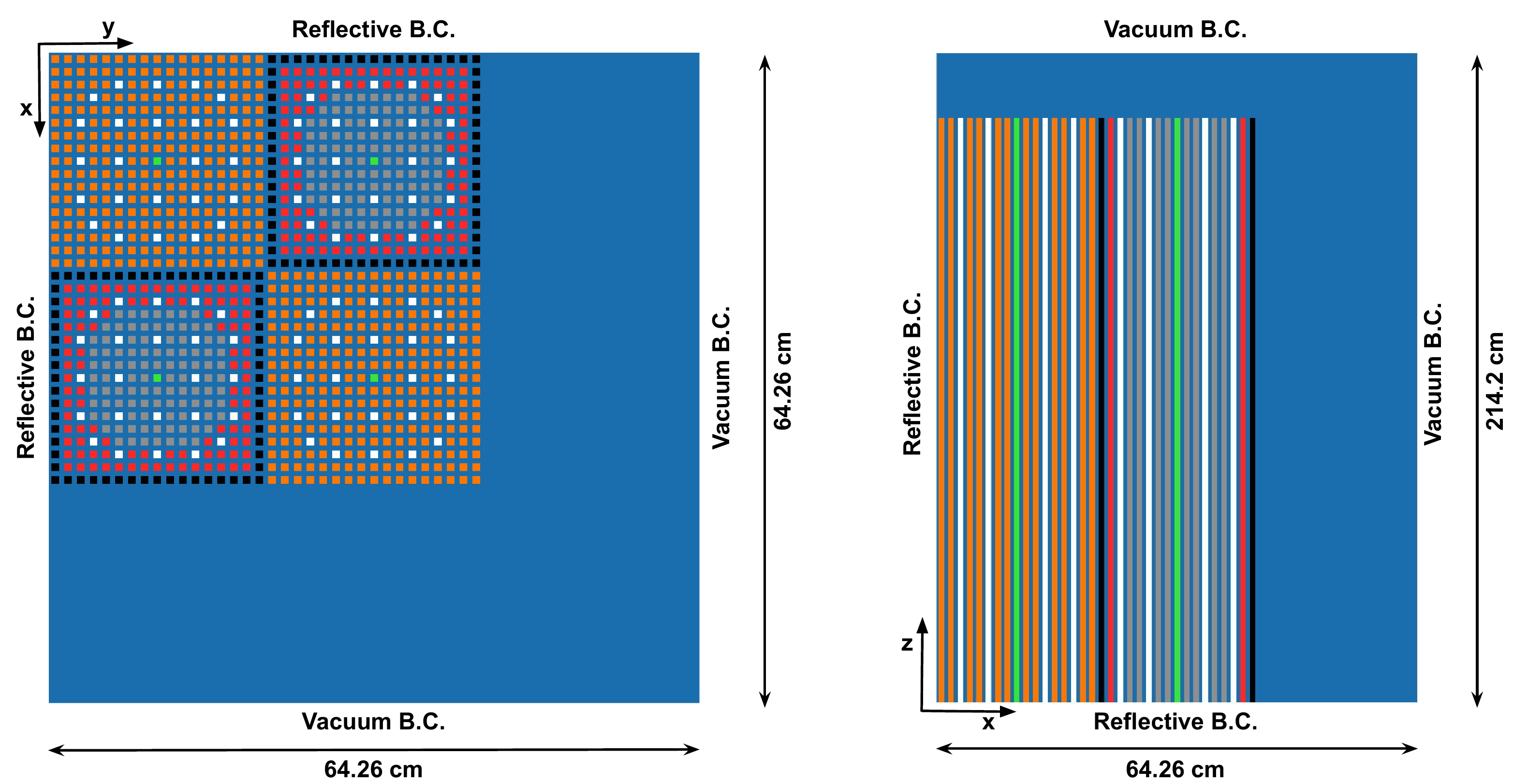}
        \caption{Geometric configuration for our modified version of the C5G7 benchmark, with boundary conditions. Each color represents a material, associated with a unique set of cross sections.}
        \label{fig:c5g7}
    \end{figure*}
    
    To test our exact, 3D multi-group regional cancellation, we used a modified version of the C5G7 international benchmark, which we have depicted in Figure~\ref{fig:c5g7} \cite{c5g7}. The C5G7 represents a small 1/8th nuclear reactor core, with four fuel assemblies and 7 energy groups, that is customarily used to assess and compare deterministic transport codes. In its original specifications, the fuel pins within the assemblies are cylindrical, with a radius of $\SI{0.54}{\centi\meter}$ \cite{c5g7}. The use of cylindrical fuel pins however makes it difficult to use arbitrarily small cancellation regions while also being able to directly calculate the minimum value of Eq.~\eqref{eq:3d_f_mg}, to be used as $\beta$. In order to facilitate cancellation, we have modified the benchmark so that the fuel pins are square in the $x$-$y$ plane, with side lengths of $\SI{0.756}{\centi\meter}$. This choice allows the fuel cells to be easily cut into an integral number of rectangular prisms to be used as cancellation regions, ensuring that no material other than fuel is present in the region. An important difference from the 1D rod system is that, in the C5G7 benchmark, the cross sections are piece-wise spatially constant, being homogeneous within a given material cell. Also, while we derived the 3D cancellation formulas for the general case of anisotropic scattering, all scattering in the C5G7 benchmark is assumed to be isotropic, simplifying Eq.~\eqref{eq:3d_f_mg} to
    \begin{equation}
        f(\bm{r}|\bm{r}_0,E_g) = \frac{\Sigma_f(\bm{r},E_g)}{4\pi\abs{\bm{r} - \bm{r}_0}^2} e^{-\Sigma_\text{smp}(\bm{r}_0,E_g)\abs{\bm{r} - \bm{r}_0}}.
        \label{eq:3d_f_mg_iso}
    \end{equation}
    Under this assumption, we have no need to keep track of the parent particle's initial direction $\hat{\bm{\Omega}}'$, as there is equal probability of scattering in any direction.
    
    The C5G7 benchmark makes use of 3 reflective boundary conditions, as shown in Figure~\ref{fig:c5g7}. One caveat of the regional cancellation method is that Eq.~\eqref{eq:3d_f_mg} is akin to a next-event point reaction rate estimator. Due to this, a trivial implementation of the method can not be used with reflective boundary conditions \cite{mcnpTheory}. Use of this method with reflective boundary conditions is only possible if special book-keeping is done to ensure a proper calculation of the expected fission density. If a particle begins at $\bm{r}_0$ and encounters a reflection before having a real collision at $\bm{r}$, then the distances which one would obtain by simply computing $\abs{\bm{r} - \bm{r}_0}$ will not represent the actual distance traveled. It is also not enough to simply store the distance traveled by the particle, as one should ideally be able to calculate the value of $f$ for any given $\bm{r}$ in the region, in order to calculate $\beta$. To ensure a proper calculation of $\abs{\bm{r} - \bm{r}_0}$, the initial position $\bm{r}_0$ of the parent particle is stored: upon each reflection, this position is transformed to the mirrored location on the other side of the plane of reflection. When doing this, the point will generally lie outside of the defined geometry for the system, but this should not pose a problem, as we only need to be able to calculate the flight distance between $\bm{r}_0$ and other points in the cancellation region. With delta tracking and NWDT, $\Sigma_\text{maj}(E_g)$ and $\Sigma_\text{smp}(E_g)$ do not depend on the position within the geometry, but only on the energy group. We therefore do not need to worry about trying to find what material this fictitious point would be located in, or the distance to the next surface.
    
    A problem of this type is admittedly a simple case, where NWDT is not strictly necessary. One is able to trivially determine the majorant cross section so that delta tracking could be used, avoiding the problem of negative weights all together. As mentioned in Sec.~\ref{sec:pid}, however, in the context of multi-physics problems which are represented with spatially continuous cross sections, determining the majorant exactly is likely to be difficult, or impossible. It would certainly be interesting to test this algorithm on a more complex problem, where delta tracking would not be possible (or would be very inefficient). It would be difficult to verify that the presented cancellation method is working with such a problem, however, as it could not easily be solved with existing methods. This is why we have chosen to examine this simpler problem.
    
    To test and evaluate cancellation and NWDT, a multi-group Monte Carlo code, called MGMC, was written to solve $k$-eigenvalue power iteration problems.
    MGMC supports general geometries using traditional surface-based descriptions of volumes.
    Either delta tracking \cite{woodcock1965techniques} or the variant of negative weighted delta tracking developed by Carter, Cashwell, and Taylor \cite{carter1972monte,legrady2017woodcock} may be selected for transport.
    Scalar flux and the fission reaction rate may be scored over a rectilinear mesh, using collision estimators\footnote{Scalar flux is defined as $\int_{4\pi}\varphi(\bm{r},\hat{\bm{\Omega}})\dd\hat{\bm{\Omega}} \text{ }$\cite{bell_glasstone}.}. Monte Carlo estimates are saved as binary Numpy files \cite{numpy2020} for easy analysis and plotting with Python. Cancellation regions are defined by a rectilinear mesh imposed on top of the problem, and can be used with both delta tracking and negative weighted delta tracking.
    Shared memory parallelism is implemented with OpenMP. Geometry, material properties, scores, cancellation, and simulation settings are all controlled with a YAML input file.
    MGMC is written in C++17, and has been made available under the CeCILL-v2.1 license \cite{mgmc}. One will also find the necessary input files to replicate our results there.
    
    All simulations were started with $10^6$ particles uniformly distributed across the four fuel assemblies, all in the first energy group, and ran for 2200 generations, with the first 200 generations being discarded to allow for fission source convergence. Delta tracking was first run (without weight cancellation) to obtain a reference multiplication factor for the system, which was found to be $k_\text{eff}=1.21912 \pm 0.00002$. When using the Carter, Cashwell, and Taylor variant of negative weighted delta tracking, the majorant cross section was used for the sampling cross section in all the energy groups except for the first; for the first energy group, $0.9\Sigma_\text{maj}(E_1)$ was used as the sampling cross section. With this choice of $\Sigma_\text{smp}$, the total cross section is underestimated in all of the fuel pins in the first energy group. Any virtual collisions which occur in these regions of the phase space will therefore result in the particle weight changing sign.
    
\begin{figure}
        \includegraphics[width=\columnwidth]{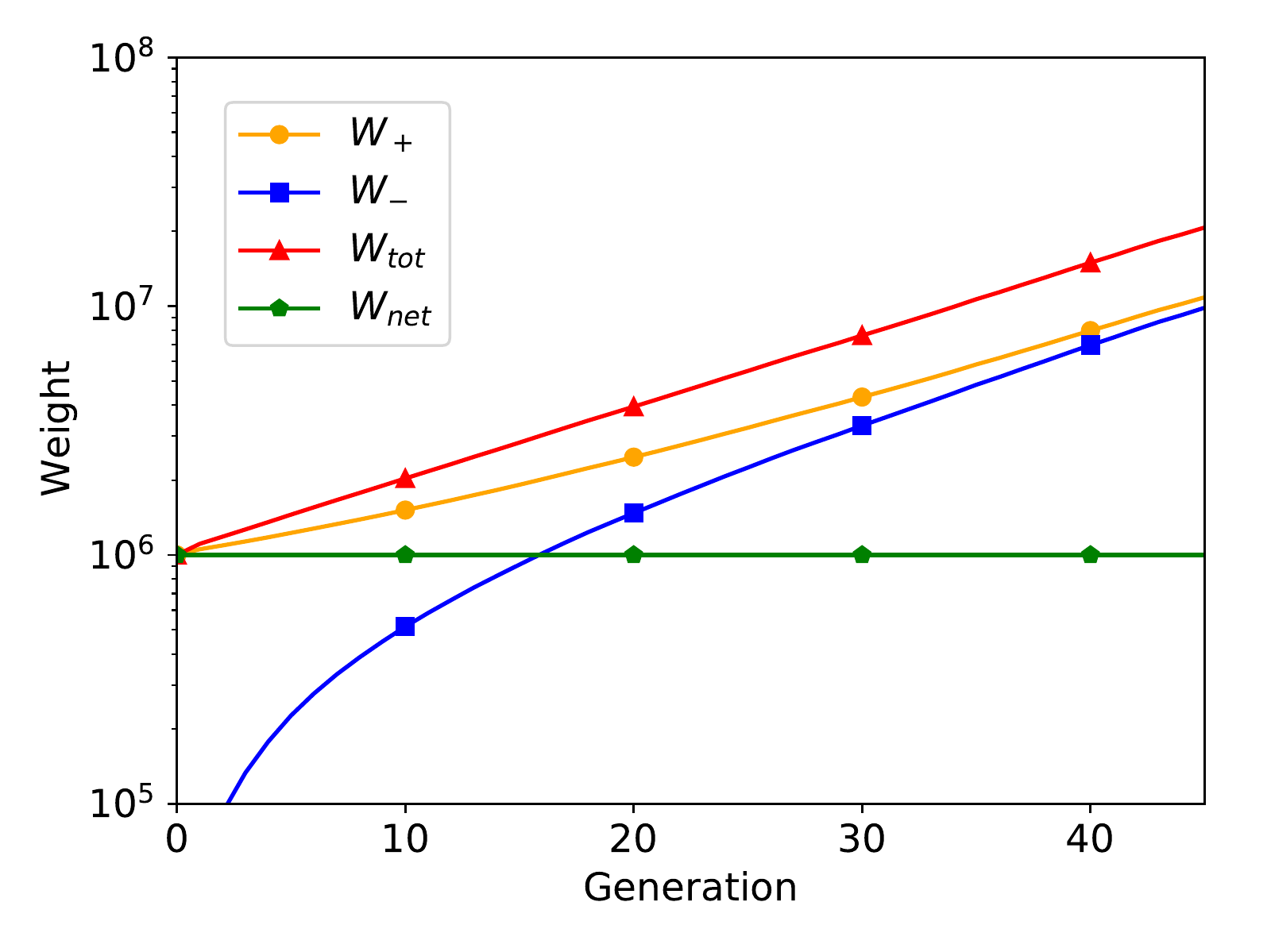}
        \caption{The positive, negative, net, and total weights in the modified C5G7 benchmark, using negative weighted delta tracking and no weight cancellation.}
        \label{fig:c5g7_no_cancel_wgt}
    \end{figure}

    Running this simulation with NWDT, and without the use of any weight cancellation, causes the particle populations to diverge and the simulation to fail. The exponential increase in total weight is depicted in Figure~\ref{fig:c5g7_no_cancel_wgt}. The behavior is nearly identical to the 1D case presented in Section~\ref{sec:pid}. Asymptotically, the negative weight will increase at the same rate as the positive weight, with the difference between the two remaining constant. The exponential increase in $W_\text{tot}$ also leads to an exponential increase in the total number of particles in the simulation, overwhelming computer memory, exactly as in the 1D case. This supports the point made by Eq~\eqref{eq:kn_gt_keff}; weight cancellation will always be necessary when attempting to perform power iteration simulations using NWDT.
    
    In order to implement regional cancellation, all fuel pins were divided into cubical cancellation regions with side lengths of $\SI{0.252}{\centi\meter}$. This mesh was chosen by trial and error, as we have found no practical way to know in advance for any system the minimum number of cancellation regions required to stabilize the simulation. Cubical cancellation regions also seemed to be more efficient than regions which had aspect ratios much larger than one, as compact region shapes lead to a higher minimum value of $f$ in the region, and therefore increases $\beta$.
    
\begin{figure}
        \includegraphics[width=\columnwidth]{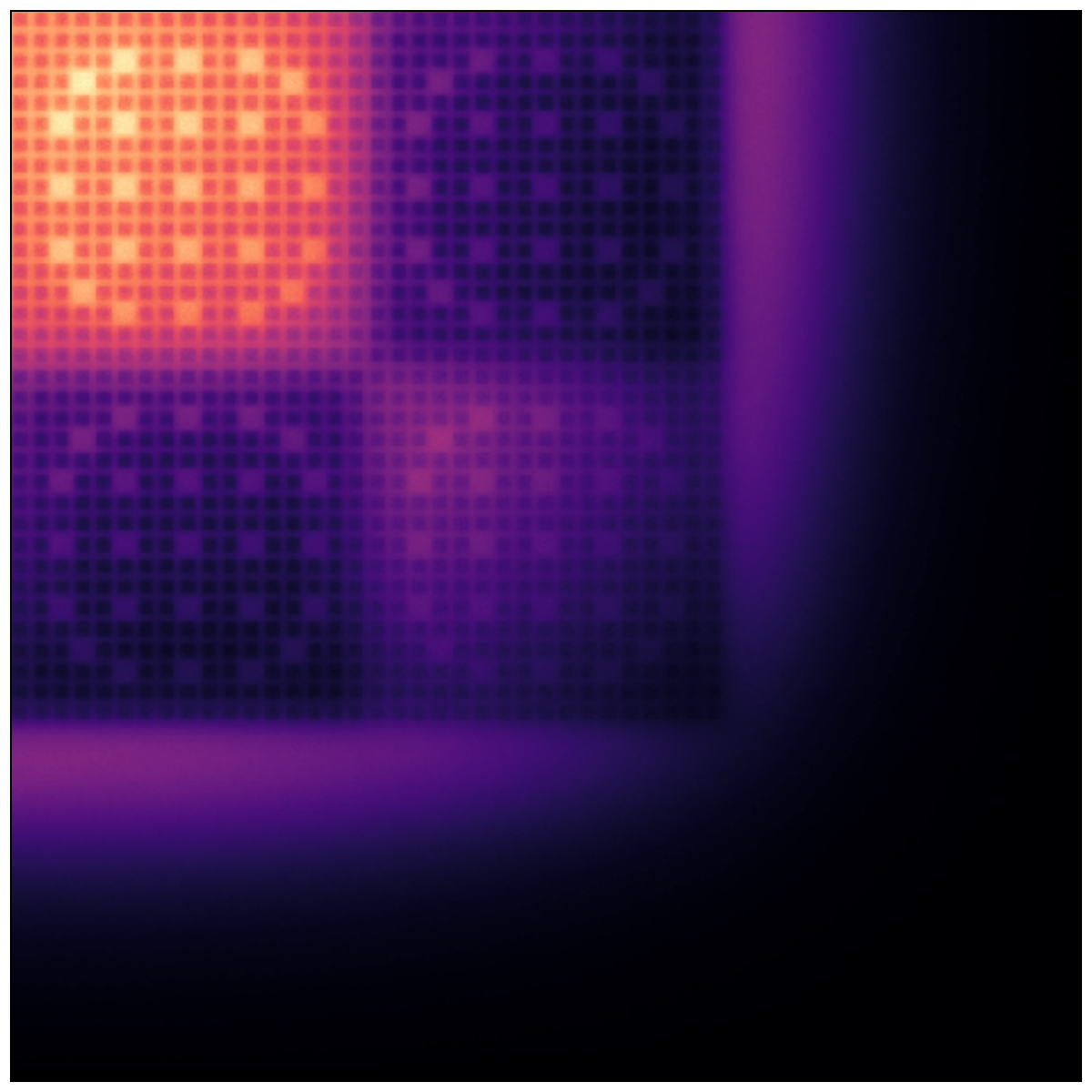}
        \caption{Scalar flux of the 7th energy group in the center axial slice of the core, obtained using negative weighted delta tracking and 3D regional cancellation.}
        \label{fig:flux}
    \end{figure}

\begin{figure}
        \includegraphics[width=\columnwidth]{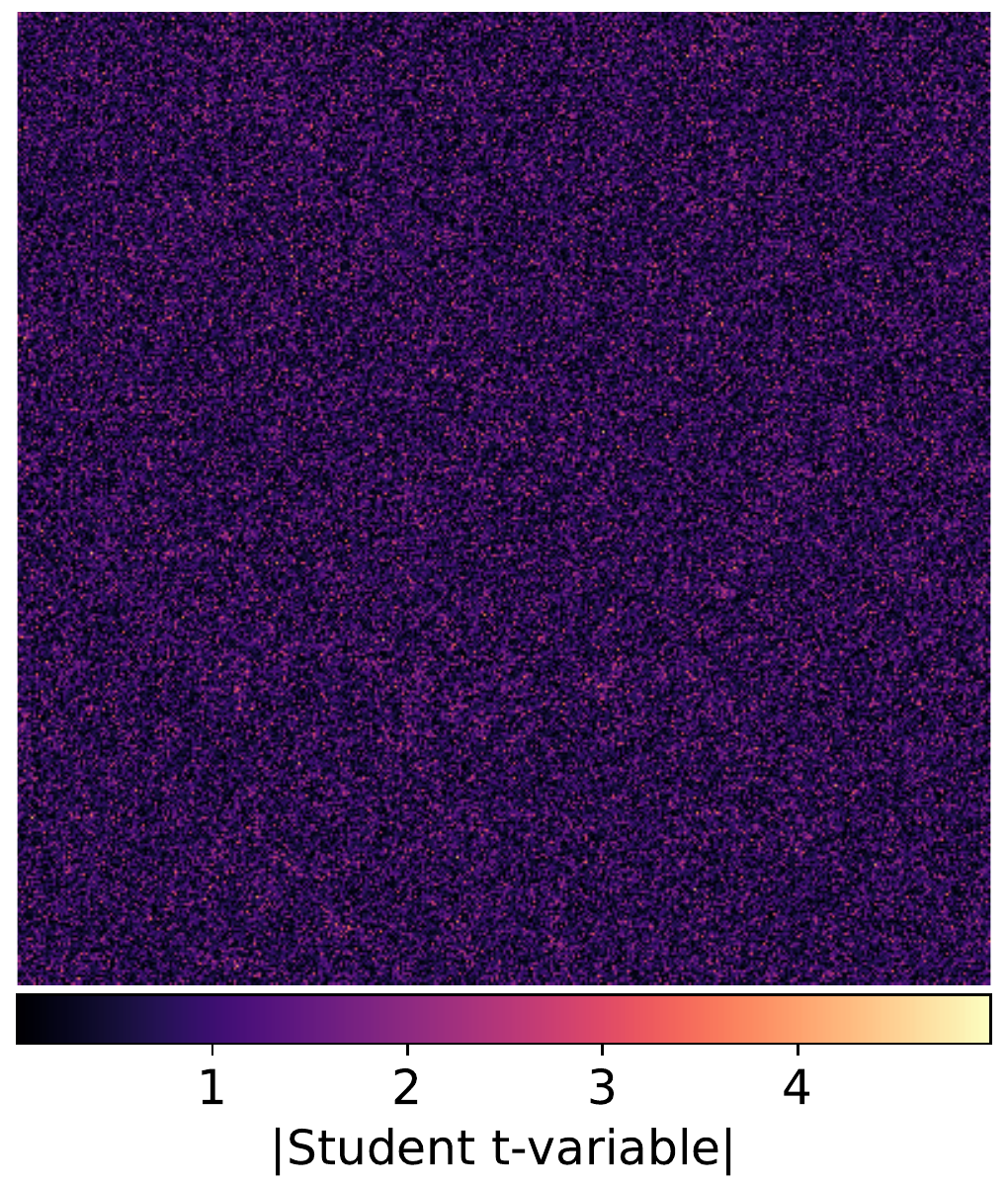}
        \caption{Absolute value of the Student $t$-statistic for the difference between the delta tracking and NWDT flux estimates. The portion displayed is for the 7th energy group, at center axial slice.}
        \label{fig:flux_err}
    \end{figure}

    Our 3D multi-group exact regional cancellation algorithm was able to stabilize the particle populations when using NWDT in conjunction with the previously outlined parameters, and an eigenvalue of $k_\text{eff}=1.21915\pm0.00005$ was obtained. This is in very good agreement with the value of $k_\text{eff}$ obtained from traditional delta tracking with no weight cancellation, as the two estimates differ by less than one standard deviation. The thermal (7th energy group) scalar flux which was tallied is shown in Figure~\ref{fig:flux}. This was compared with the flux tally from the delta tracking simulation by performing a Student $t$-test in each mesh cell \cite{stats_book}. The plot of the absolute value of the Student $t$-statistic is provided in Figure~\ref{fig:flux_err}.
    \begin{figure}
        \includegraphics[width=\columnwidth]{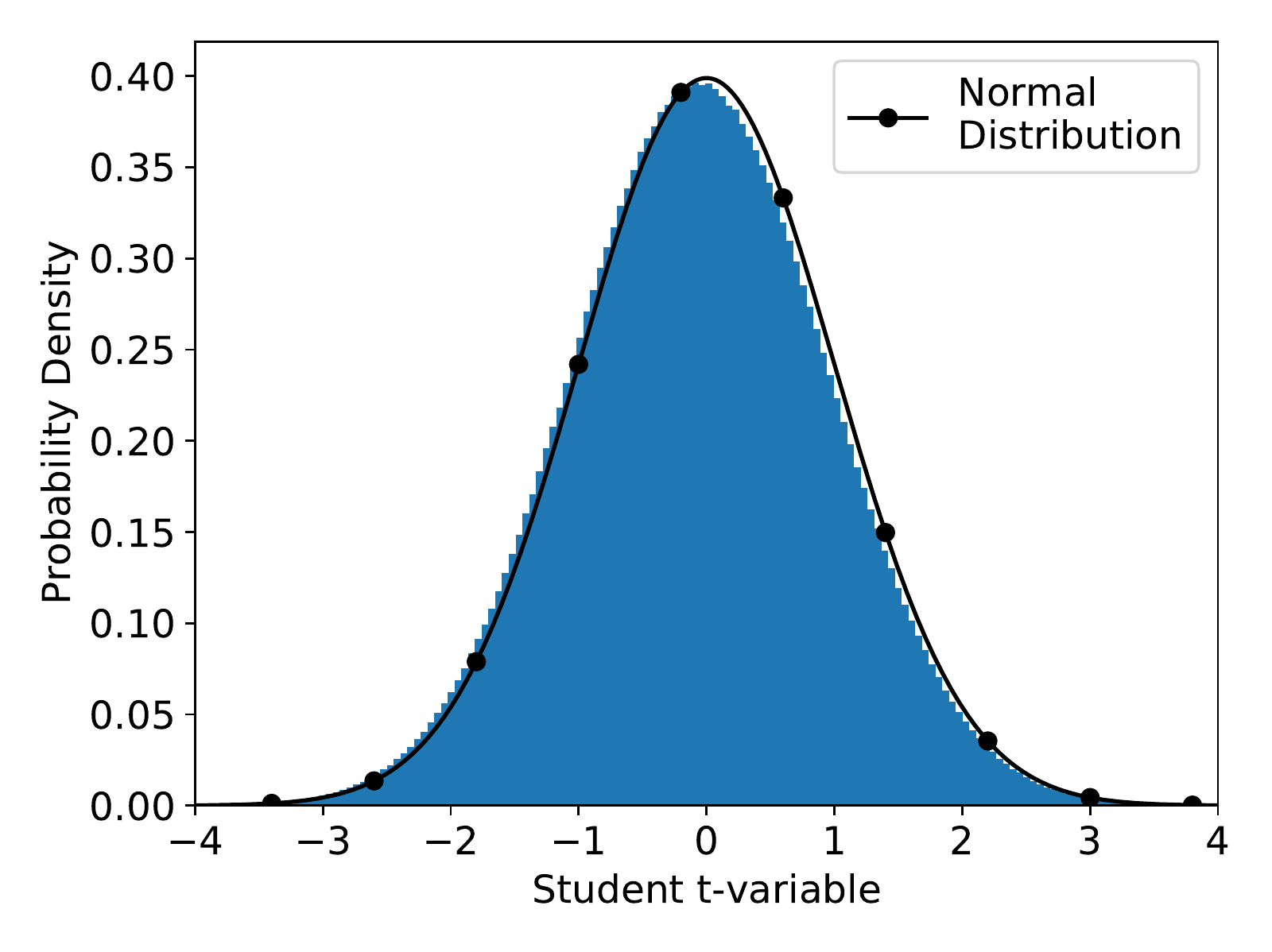}
        \caption{Histogram of the Student $t$-statistic for the difference between the flux estimates obtained with delta tracking and NWDT. The curve plotted on top is the expected normal distribution.}
        \label{fig:t_dist}
    \end{figure}
    It can be seen there that the two flux estimates are also in good agreement with one another, with no apparent spatial dependence in their difference.
    
    These two maps only show a small portion of the examined phase space. For a more thorough comparison, a histogram of the distribution of the Student $t$-statistic for the NWDT flux and the delta tracking flux is given in Figure~\ref{fig:t_dist} \cite{stats_book}. Large portions of the flux tally mesh have average values of zero, due to the large water reflectors. These elements were removed before producing Figure~\ref{fig:t_dist}. Elements which had a relative error greater than 20\% (in either the NWDT or the delta tracking score) were also removed, as the Student $t$-test is only applicable on normally distributed variables; therefore, we need to ensure that sufficient statistics are accumulated, so that the central limit theorem may apply. The 20\% cut-off is admittedly somewhat arbitrary, but it is likely that bins with such a high relative error are not normally distributed. It can be seen that the histogram has very good agreement with the theoretical distribution, which we assume to be normal given the large number of degrees of freedom for the comparison.
    
\begin{figure}
        \includegraphics[width=\columnwidth]{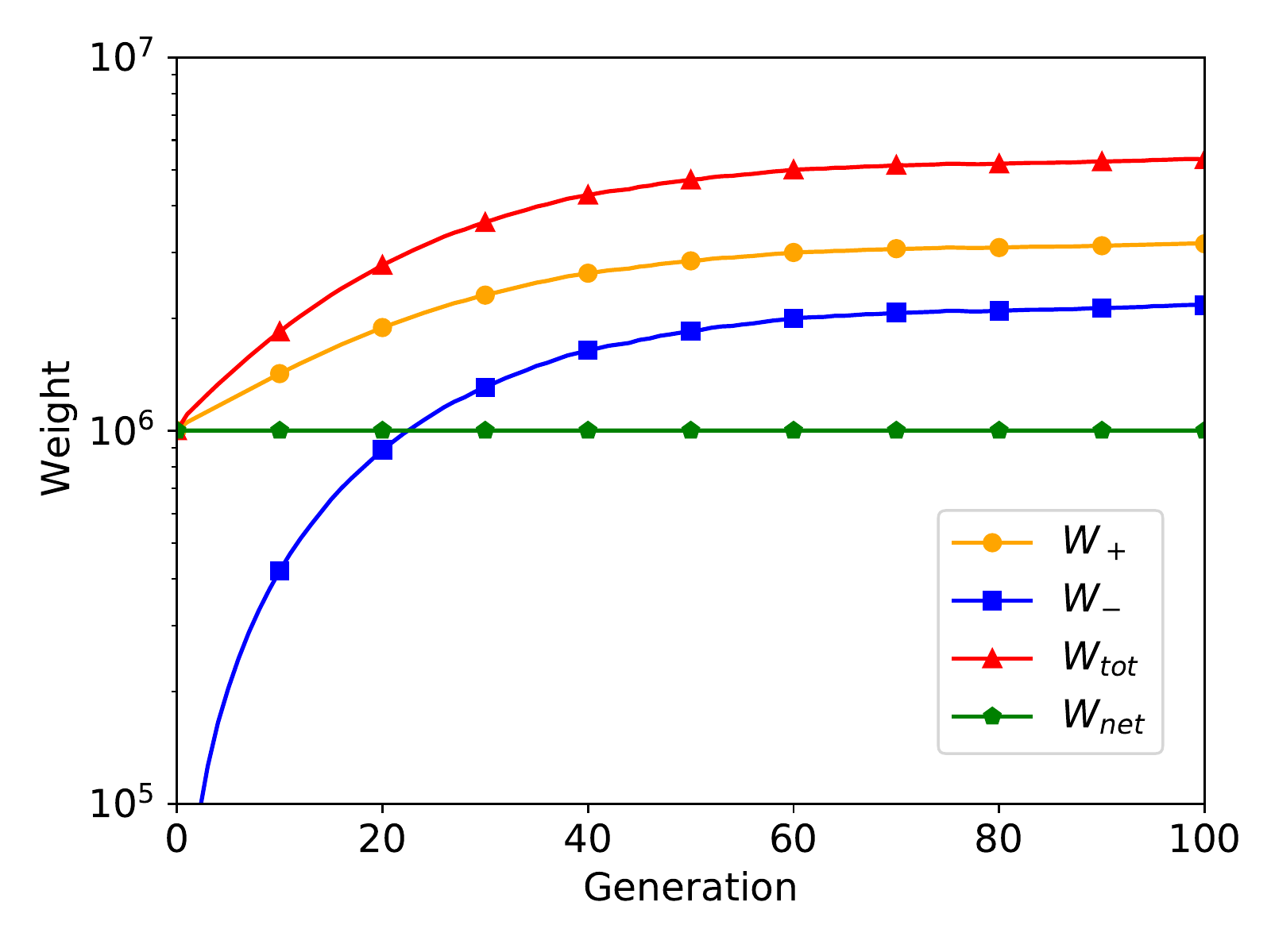}
        \caption{The positive, negative, net, and total weights in the modified C5G7 benchmark, using negative weighted delta tracking with exact 3D regional weight cancellation.}
        \label{fig:c5g7_wgt}
    \end{figure}

    Finally, Figure~\ref{fig:c5g7_wgt} shows the behaviour of the weights in the system for the first 100 generations in the simulation. As with the 1D case, there is initially no negative weight. This increases rapidly in the first 10-20 generations, before beginning to level out. By the time 80 generations have passed, the weights have reached their equilibrium values. This is much longer than the 1D case, which only required 2--3 generations before the weights reached equilibrium. The longer time to convergence is attributed to the C5G7 benchmark having a higher dominance ratio than the 1D case.
    
    % I really would like to see more of this computational framework in action.
    % After all the theoretical buildup, I felt like it would be nice to get some more
    % computational results that showcase the results “experimentally.” In particular,
    % it would be really nice to talk about (a) what happens without negative weight
    % cancellation, so as to demonstrate the importance of the new algorithm, and (b) a
    % demonstration of this framework in some case where regular delta tracking is not
    % possible, to demonstrate the utility of the method in a currently intractable
    % situation.
    
    \section{Conclusions}
    \label{sec:conclusions}
    
    Over the course of this work, we have presented a previously undocumented population control problem which arises when attempting to run $k$-eigenvalue power iteration Monte Carlo simulations with both negative and positive weights in negative weighted delta tracking \cite{legrady2017woodcock}. Modeling the transport process through a set of coupled Boltzmann transport equations for negative and positive particles, we were able to show that the power iteration technique applied to negative weighted delta tracking will always fail to converge, as the sought physical fundamental eigenvalue is not the dominant eigenvalue of the system. Instead, a fictitious eigenvalue, associated with a system with a lower amount of absorption, is now dominant. We have formally developed a cancellation operator, demonstrating theoretically, and with a deterministic model, how particle weight cancellation can suppress the fictitious eigenstate and restore convergence of power iteration to the physical fundamental eigenvalue.
    
    To demonstrate weight cancellation in a Monte Carlo context, we have implemented the exact 1D region cancellation algorithm of Booth and Gubernatis \cite{booth2010exact} in a 1D rod model. Weight cancellation did stabilize the particle populations through the fission generations, allowing the simulation to finish. As the method outlined by Booth and Gubernatis is only valid in 1D single-speed problems, we have developed an exact 3D regional cancellation method. We tested our 3D algorithm on a modified version of the C5G7 reactor-physics benchmark, using negative weighted delta tracking. Cancellation stabilized the particle populations, and also resulted in estimates for the multiplication factor and flux which were in agreement with the reference results obtained through delta tracking. Our exact 3D multi-group regional cancellation could potentially be useful for other applications in neutron transport as well. One such case is the use of Monte Carlo methods to obtain higher harmonics of the $k$-eigenvalue equation \cite{booth2003computing}. Our cancellation algorithm could also potentially prove useful in transport problems involving complex particle weights, such as the search for critical buckling or the solution of the neutron noise equations \cite{yamamoto2012buckling, yamamoto2013}.
    
    Several questions remain to be settled in regard to the methodology of 3D cancellation. First, in this work we have required that a cancellation region consist of only one material. Being able to have multiple materials in a cancellation region would make it easier to perform cancellation on different geometric forms (cylindrical pins). Second, extension of this algorithm to continuous energy will require the application of new techniques, because in continuous-energy transport the fission spectrum depends on the incoming neutron energy. Finally, the efficient choice of $\beta$ should be evaluated. While we have chosen to use the minimum value of the expected fission density in the region for $\beta$, Booth and Gubernatis made clear that any value of $\beta$ results in an unbiased estimate of the fission density. However, the variance of the weights of the fission particles clearly depends on $\beta$. It is desirable to characterize which choices of $\beta$ result in more efficient cancellation. All of these issues will require further investigation to improve the algorithm, and to probe its possible extension to continuous-energy problems.
    
    \appendix*
    \section{Derivation of coupled Boltzmann transport equations}
    
    We provide here the derivation of the coupled system of Boltzmann equations, where the populations of positive and negative particles are treated separately and are assumed to be governed by the rules of negative weighted delta tracking (Alg.~\ref{alg:ndwt}) \cite{legrady2017woodcock}. The quantities that we wish to describe are the angular fluxes of positive and negative particles. These quantities are not physical observables, but they are useful to characterize the behaviour of the random walk. Consider any one of the usual Monte Carlo estimators for the (physical) angular flux; the unbiasedness condition requires any such estimator to be proportional to the particle weight $w$. Replacing the particle weight $w$ in the estimator with
    \begin{equation}
      w_+=\begin{cases}w& w>0\\0 &w\leq 0\end{cases}
    \end{equation}
    results in a modified estimator; we define the angular flux of positive particles $\varphi_+$ to be the expected value of the modified estimator. Likewise, replacing the particle weight with
    \begin{equation}
      w_-=\begin{cases}0& w>0\\-w &w\leq 0\end{cases}
    \end{equation}
    results in a modified estimator, whose expected value is defined to be the angular flux of negative particles, $\varphi_-$.
    
    A few properties are worth stressing. First, if all particle weights are positive, then $\varphi_+=\varphi$ (the physical angular flux) and $\varphi_-=0$. Second, since $w_+-w_-=w$, then $\varphi_+-\varphi_-=\varphi$. Finally, since $w_+\geq 0$ and $w_-\geq 0$, then both $\varphi_+$ and $\varphi_-$ are always non-negative.
    
    With these postulations, we may commence our derivation of the angular flux of positive particles. We shall obey the rules for negative weighted delta tracking \cite{legrady2017woodcock}, presented in Alg.~\ref{alg:ndwt}. The Boltzmann equation for neutron transport is canonically written in the form of a balance equation, where losses = gains. These losses and gains refer to the change in the neutron flux at the phase space point $(\bm{r},\hat{\bm{\Omega}})$. This can be referenced in Eq.~\eqref{eq:boltzmann}, where the left hand side (LHS) of the equation represents losses, while the right hand side (RHS) represents gains. More detail as to the reasoning and derivation behind particular terms can be found in \textit{Nuclear Reactor Theory}, by Bell and Glasstone \cite{bell_glasstone}.
    
    Starting with the LHS, we first consider losses due to particle streaming, $L_S$. This term is identical to that found in Eq.~\eqref{eq:boltzmann}, only replacing the physical angular flux with the angular flux of positive particles:
    \begin{eqnarray}
        L_S = \hat{\bm{\Omega}}\cdot\nabla\varphi_+(\bm{r},\hat{\bm{\Omega}}).
    \end{eqnarray}
    Next, we consider losses due to collisions, $L_C$. In the case of NWDT, it is possible to be removed from the phase space point by having a collision at $\bm{r}$ (real or virtual), which occurs with a cross section $\Sigma_\text{smp}$. We therefore consider the losses due to collisions with the term
    \begin{eqnarray}
        L_C = \Sigma_\text{smp}(\bm{r})\varphi_+(\bm{r},\hat{\bm{\Omega}}).
    \end{eqnarray}
    In the event of a real collision, the particle direction will change, so there is indeed a removal from the phase space point. If a virtual collision occurs and the positive particle changes sign to be a negative particle, this is also a loss. If a virtual collision does not result in a positive particle becoming negative, then it is not actually a loss. This case will be treated on the RHS by adding a gain term for virtual collisions which do not result in a sign flip. The only possible sources for loss is then $L_S$ and $L_C$, and we then have a LHS of
    \begin{equation}
        L_S + L_C = \hat{\bm{\Omega}}\cdot\nabla\varphi_+(\bm{r},\hat{\bm{\Omega}}) + \Sigma_\text{smp}(\bm{r})\varphi_+(\bm{r},\hat{\bm{\Omega}}).
    \end{equation}
    
    Next, we may consider gains at the phase space point $(\bm{r},\hat{\bm{\Omega}})$. Gains may only come from particles which have had either a real or virtual collision, allowing them to enter the phase space point in question. First, we shall consider the gains from real collisions, $G_R$. A real collision occurs with probability $q(\bm{r})$, and when this happens, a weight modification also occurs, which is a multiplication of the factor $\Sigma_t(\bm{r})/(q(\bm{r})\Sigma_\text{smp}(\bm{r}))$:
    \begin{eqnarray}
        G_R = q(\bm{r})\frac{\Sigma_t(\bm{r})}{q(\bm{r})\Sigma_\text{smp}(\bm{r})}R,
        \label{eq:real_gains}
    \end{eqnarray}
    here $R$ is the sum of the gains of all real collision channels. We have two possible reaction channels which contribute to gains at $(\bm{r},\hat{\bm{\Omega}})$: particles having a collision at $\bm{r}$ and scattering from direction $\hat{\bm{\Omega}}'$ into direction $\hat{\bm{\Omega}}$ ($R_S$), and particles having a collision at $\bm{r}$, inducing a fission particle which is born traveling in direction $\hat{\bm{\Omega}}$ ($R_F$). Treating the scattering term, we must sum over all possible incoming directions, resulting in
    \begin{equation}
        R_S = \frac{\Sigma_\text{smp}(\bm{r})}{\Sigma_t(\bm{r})}\int_{4\pi}\Sigma_s(\bm{r},\hat{\bm{\Omega}}'\rightarrow\hat{\bm{\Omega}})\varphi_+(\bm{r},\hat{\bm{\Omega}}')\dd\hat{\bm{\Omega}}'.
        \label{eq:real_scat}
    \end{equation}
    We must multiply by the ratio $\Sigma_\text{smp}/\Sigma_t$ to account for the fact that we are conditioning on the collision being real. We now perform a similar operation for the fission channel, with the assumption that the distribution of the direction of fission particles is isotropic:
    \begin{eqnarray}
        R_F = \frac{\Sigma_\text{smp}(\bm{r})}{\Sigma_t(\bm{r})}\frac{\nu(\bm{r})}{4\pi k}\int_{4\pi}\Sigma_f(\bm{r})\varphi_+(\bm{r},\hat{\bm{\Omega}}')\dd\hat{\bm{\Omega}}'.
        \label{eq:real_fis}
    \end{eqnarray}
    These being the only two real collision channels ($R = R_S + R_F$), we may combine the definitions of $R_S$, $R_F$, and $G_R$ to obtain the result presented in Eq.~\eqref{eq:gains_real}, where we have employed our previous definitions for $\mathcal{S}$ and $\mathcal{F}$ from Eq.~\eqref{eq:S_deff} and Eq.~\eqref{eq:F_deff}.
    \begin{widetext}
    \begin{eqnarray}
        G_R =& q(\bm{r})\frac{\Sigma_t(\bm{r})}{q(\bm{r})\Sigma_\text{smp}(\bm{r})} \bigg[ \frac{\Sigma_\text{smp}(\bm{r})}{\Sigma_t(\bm{r})}\int_{4\pi}\Sigma_s(\bm{r},\hat{\bm{\Omega}}'\rightarrow\hat{\bm{\Omega}})\varphi_+(\bm{r},\hat{\bm{\Omega}}')\dd\hat{\bm{\Omega}}' +  \frac{\Sigma_\text{smp}(\bm{r})}{\Sigma_t(\bm{r})}\frac{\nu(\bm{r})}{4\pi k}\int_{4\pi}\Sigma_f(\bm{r})\varphi_+(\bm{r},\hat{\bm{\Omega}}')\dd\hat{\bm{\Omega}}'\bigg] \nonumber\\
        =& \int_{4\pi}\Sigma_s(\bm{r},\hat{\bm{\Omega}}'\rightarrow\hat{\bm{\Omega}})\varphi_+(\bm{r},\hat{\bm{\Omega}}')\dd\hat{\bm{\Omega}}' + \frac{\nu(\bm{r})}{4\pi k}\int_{4\pi}\Sigma_f(\bm{r})\varphi_+(\bm{r},\hat{\bm{\Omega}}')\dd\hat{\bm{\Omega}}'\nonumber\\
        =& \mathcal{S}\varphi_+ + \frac{1}{k}\mathcal{F}\varphi_+
        \label{eq:gains_real}
    \end{eqnarray}
    \end{widetext}
    
    All that remains are gains from virtual collisions $G_V$. The probability of a virtual collision is $1-q(\bm{r})$, and is accompanied by a weight modification of
    \begin{eqnarray}
        \frac{\abs{1-\frac{\Sigma_t(\bm{r})}{\Sigma_\text{smp}(\bm{r})}}}{1-q(\bm{r})}.
    \end{eqnarray}
    We take the absolute value here, as our particle weights may never become negative. Instead, a change in sign is modeled by the transfer of a particle from the positive population (corresponding to $\varphi_+$) to the negative population (corresponding to $\varphi_-$). Our virtual gains are
    \begin{eqnarray}
        G_V = (1-q(\bm{r}))\frac{\abs{1-\frac{\Sigma_t(\bm{r})}{\Sigma_\text{smp}(\bm{r})}}}{1-q(\bm{r})}V,
    \end{eqnarray}
    $V$ being the sum of gains due to virtual collisions. The first channel for virtual collisions is the previously mentioned case of a positive particle having a virtual collision, and remaining positive ($V_{+}$). This only occurs when $\Sigma_\text{smp}(\bm{r}) \ge \Sigma_t(\bm{r})$, and can be modeled with the Heaviside function
    \begin{eqnarray}
        \Theta(x) = \begin{cases} 1 & x \ge 0 \\ 0 & x < 0 \end{cases}.
    \end{eqnarray}
    Our gains from virtual collisions of positive particles is then
    \begin{eqnarray}
        V_+ = \Sigma_\text{smp}(\bm{r})\Theta\big(\Sigma_\text{smp}(\bm{r}) - \Sigma_t(\bm{r})\big)\varphi_+(\bm{r},\hat{\bm{\Omega}}).
    \end{eqnarray}
    
    In the event that $\Sigma_\text{smp}(\bm{r}) < \Sigma_t(\bm{r})$, negative particles will flip sign, joining the positive particles ($V_-$). This is modeled in a similar manner, simply flipping the argument in the Heaviside function, and replacing the positive angular flux with the negative angular flux:
    \begin{eqnarray}
        V_- = \Sigma_\text{smp}(\bm{r})\Theta\big(\Sigma_t(\bm{r}) - \Sigma_\text{smp}(\bm{r})\big)\varphi_-(\bm{r},\hat{\bm{\Omega}}).
    \end{eqnarray}
    With these terms defined, we may combine our definitions of $G_V$ and $V=V_+ + V_-$ to obtain the gains from virtual collisions, presented in Eq.~\eqref{eq:gains_virt}, having used the function from Eq.~\eqref{eq:diode}.
    
    All loss and gain terms have now been defined, leaving us with the final form of the equation, describing the angular flux of the positive particles in Eq.~\eqref{eq:boltz_pos}. By symmetry, the equation for negative particles must have the exact same form, and is given in Eq.~\eqref{eq:boltz_neg}. Regardless of whether a particle is negative or positive, it is transported in the same manner, and by the same rules. These two equations are of course the form presented in Eq.~\eqref{eq:cpld}, and demonstrate the coupling relationship between positive and negative particles.
    \begin{widetext}
    \begin{eqnarray}
        G_V = (1-q(\bm{r}))\frac{\abs{1-\frac{\Sigma_t(\bm{r})}{\Sigma_\text{smp}(\bm{r})}}}{1-q(\bm{r})}\Sigma_\text{smp}(\bm{r})\bigg[\Theta\big(\Sigma_\text{smp}(\bm{r}) - \Sigma_t(\bm{r})\big)\varphi_+(\bm{r},\hat{\bm{\Omega}}) + \Theta\big(\Sigma_t(\bm{r}) - \Sigma_\text{smp}(\bm{r})\big)\varphi_-(\bm{r},\hat{\bm{\Omega}})\bigg] \nonumber\\
        = \Theta\big(\Sigma_\text{smp}(\bm{r}) - \Sigma_t(\bm{r})\big)\big[\Sigma_\text{smp}(\bm{r}) - \Sigma_t(\bm{r})\big]\varphi_+(\bm{r},\hat{\bm{\Omega}}) + \Theta\big(\Sigma_t(\bm{r}) - \Sigma_\text{smp}(\bm{r})\big)\big[\Sigma_t(\bm{r}) - \Sigma_\text{smp}(\bm{r})\big]\varphi_-(\bm{r},\hat{\bm{\Omega}}) \nonumber\\
        = \Delta\big(\Sigma_\text{smp}(\bm{r}) - \Sigma_t(\bm{r})\big)\varphi_+(\bm{r},\hat{\bm{\Omega}}) + \Delta\big(\Sigma_t(\bm{r}) - \Sigma_\text{smp}(\bm{r})\big)\varphi_-(\bm{r},\hat{\bm{\Omega}})
        \label{eq:gains_virt}
    \end{eqnarray}
    \begin{eqnarray}
        \hat{\bm{\Omega}}\cdot\nabla\varphi_+ + \Sigma_\text{smp}\varphi_+ = \mathcal{S}\varphi_+ + \frac{1}{k}\mathcal{F}\varphi_+ + \Delta\big(\Sigma_\text{smp} - \Sigma_t\big)\varphi_+ + \Delta\big(\Sigma_t - \Sigma_\text{smp}\big)\varphi_- \label{eq:boltz_pos} \\ \hat{\bm{\Omega}}\cdot\nabla\varphi_- + \Sigma_\text{smp}\varphi_- = \mathcal{S}\varphi_- + \frac{1}{k}\mathcal{F}\varphi_- + \Delta\big(\Sigma_\text{smp} - \Sigma_t\big)\varphi_- + \Delta\big(\Sigma_t - \Sigma_\text{smp}\big)\varphi_+
        \text.
        \label{eq:boltz_neg}
    \end{eqnarray}
    \end{widetext}
    
    % To be replaced with .bbl for arXiv
    %\bibliographystyle{apsrev4-2}
    %\bibliography{references}
    \input{main.bbl}

\end{document}

%% file: main.bbl
%apsrev4-2.bst 2019-01-14 (MD) hand-edited version of apsrev4-1.bst
%Control: key (0)
%Control: author (72) initials jnrlst
%Control: editor formatted (1) identically to author
%Control: production of article title (-1) disabled
%Control: page (0) single
%Control: year (1) truncated
%Control: production of eprint (0) enabled
%